\documentclass[english,aps,superscriptaddress,preprintnumbers,reprint,footinbib,amsmath,amssymb,prb,longbibliography]{revtex4-2}
\usepackage[latin9]{inputenc}
\usepackage{xcolor}
\usepackage{babel}
\usepackage{amsmath}
\usepackage{amssymb}
\usepackage{graphicx}
\usepackage{esint}
\usepackage[]
 {hyperref}

\makeatletter
\usepackage{babel}

\usepackage{color}

\@ifundefined{textcolor}{}{%
 \definecolor{BLACK}{gray}{0}
 \definecolor{WHITE}{gray}{1}
 \definecolor{RED}{rgb}{1,0,0}
 \definecolor{GREEN}{rgb}{0,1,0}
 \definecolor{BLUE}{rgb}{0,0,1}
 \definecolor{CYAN}{cmyk}{1,0,0,0}
 \definecolor{MAGENTA}{cmyk}{0,1,0,0}
 \definecolor{YELLOW}{cmyk}{0,0,1,0}
}

\usepackage{aecompl}

\usepackage{epsfig}\usepackage{dcolumn}\usepackage{bm}

\usepackage{hyperref}
\hypersetup{
    colorlinks=true,
    linkcolor=blue,
    urlcolor=blue,
    citecolor=blue
        }

\makeatother

\begin{document}
\title{Revisiting the Poor Man's Majoranas: \\
The Spin-Exchange Induced Spillover Effect}
\author{J.E. Sanches}
\email[corresponding author:]{jose.sanches@unesp.br}

\affiliation{São Paulo State University (Unesp), School of Engineering, Department
of Physics and Chemistry, 15385-007, Ilha Solteira-SP, Brazil}
\author{T.M. Sobreira }
\affiliation{Instituto de Física, Universidade Federal Fluminense, 24210-340, Niterói,
Rio de Janeiro, Brazil}
\author{L.S. Ricco}
\affiliation{Science Institute, University of Iceland, Dunhagi-3, IS-107, Reykjavik,
Iceland}
\affiliation{Institute of Experimental Physics, Faculty of Physics, University
of Warsaw, ul. Pasteura 5, PL-02-093 Warsaw, Poland}
\author{M.S. Figueira}
\affiliation{Instituto de Física, Universidade Federal Fluminense, 24210-340, Niterói,
Rio de Janeiro, Brazil}
\author{A.C. Seridonio}
\email[corresponding author:]{antonio.seridonio@unesp.br}

\affiliation{São Paulo State University (Unesp), School of Engineering, Department
of Physics and Chemistry, 15385-007, Ilha Solteira-SP, Brazil}
\begin{abstract}
We give a review on Poor Man's Majorana (PMM) modes, which are theoretically
established in the minimal Kitaev chain implementation consisting
of two grounded, spinless quantum dots (QDs) operating at the \textit{sweet
spot} condition, where electron cotunneling and crossed Andreev reflection
amplitudes achieve precise balance. Particularly, we systematically
review, within the Green's functions theoretical framework, the PMM
hybridization dynamics under spin-exchange perturbations proposed
by some of us in \href{https://iopscience.iop.org/article/10.1088/1361-648X/adce6a}{J. Phys.: Condens. Matter 37, 205601 (2025)},
which demonstrates a characteristic spatial delocalization when subjected
to an exchange coupling $J$ mediated by a quantum spin $S$. This
spin-exchange induced PMM spillover effect provides a spectroscopic
protocol for determining the quantum statistics of $S$ through the
emergent multi-level structure in the proximal QD's density of states.
Our principal theoretical result establishes that the exchange interaction
generates $2S+2$ ($2S+1$) satellite states symmetrically distributed
about the zero-bias anomaly, serving as a definitive signature of
bosonic (fermionic) spin statistics. As novelty, we demonstrate that
multi-terminal environmental coupling induces significant suppression
of the spin-exchange spillover mechanism. Under constrained variations
of $J$, this effectively localizes the perturbed PMM within its host
QD, preventing spatial hybridization with adjacent site. The absence
of topological protection in this minimal Kitaev realization is strategically
leveraged to: (i) Develop a novel spectroscopic technique for quantum
spin characterization through PMM hybridization signatures;{{}
}(ii) Propose the \textquotedblleft environmentally induced protection\textquotedblright ,
an engineered dissipative spectral stabilization for PMMs against
exchange fluctuations in multi-terminal architectures.
\end{abstract}
\maketitle

\section{Introduction}

The theoretical framework for Majorana fermions originated with Ettore
Majorana's 1937 solution to the Dirac equation, predicting self-conjugate
fermionic excitations~\cite{Majorana-1937}. {While elementary Majorana
particles remain undetected in high-energy physics, their condensed
matter analogues, Majorana bound states (MBSs), emerge as topologically
protected zero-energy modes in superconducting systems~\cite{Kitaev-2001,Marra-2022,Flensberg-2021,Klinovaja-2021,Alicea-2012,Flensberg-2012}.}
{Kitaev's paradigmatic model of a 1D spinless $p$-wave superconductor,
the so-called Kitaev wire, establishes the theoretical foundation
for MBSs~\cite{Kitaev-2001}, demonstrating their non-local correlations and non-Abelian statistics
when a two-dimensional network of Kitaev wires is built (see details
in Sec. III of Ref.\cite{Marra-2022}).}
{The potential for topological quantum computation has driven extensive
experimental efforts, though definitive verification remains challenging
due to competing trivial states \cite{AndreevVersusMajorana,Ricco_2019}.
Recent advances have focused on developing robust platforms for MBS
characterization and manipulation. We highlight modern experimental
implementations, which require three key ingredients, such as strong
Rashba-type spin-orbit coupling (SOC), substantial Zeeman splitting
and proximity-induced $s$-wave superconductivity~\cite{Beenakker-2013,Tewari-2013,Nayak-2015,Aguado-2017,YoichiAndo-2017,Yuval-Oreg-2019}.
Therefore, two main architectures have emerged represented by semiconductor
nanowires (InSb, InAs) with induced superconductivity~\cite{DasSarma-2010,Oppen-2010,Mourik-2012,C.Marcus-2016,C.Marcus-2017,Nonlocality_Majorana,Baranger-2011}
and magnetic atom chains (Fe, Mn) on superconducting surfaces~\cite{Beenakker-2011,Oppen-2013,Klinovaja-2013,J.Franke-2015,Pawlak_2016,Roland1,Roland2,Yazdani-2014,Yazdani-2013,Yazdani-2017,Yazdani-2021}.
Nevertheless, critical challenges persist in distinguishing genuine
MBSs from trivial zero-bias peaks arising from disorder, Andreev bound
states (ABSs) or other localized states, which unfortunately emulate
a true MBS hallmark ~\cite{AndreevVersusMajorana,Ricco_2019}.}

{In light of fabrication challenges for extended topological systems,
quantum dot (QD) arrays have emerged as experimentally viable platforms
for implementing truncated Kitaev-like Hamiltonians~\cite{Dephasing,Kouwenhoven2023,MinimalModel2,MinimalModel5,MinimalModel7,MinimalModel3,MinimalModel4,MinimalModel6}.}
{This paradigm builds upon foundational work by Sau and Das Sarma~\cite{sau2012realizing}
and Leijnse-Flensberg~\cite{Flensberg_2012(Poor)}, with the latter
introducing the concept of ``Poor Man's Majoranas'' (PMMs) - Majorana-like
excitations in QD chains exhibiting partial but non-topological protection~\cite{Sanches_2025,NonHermitian1,Poor2,Poor5,Poor6,Poor11}.
Recent quantum transport measurements in coupled two- and three-QD
configurations have revealed signatures consistent with PMM formation~\cite{Dephasing,Kouwenhoven2023,MinimalModel2,MinimalModel3,MinimalModel4,MinimalModel5,MinimalModel7,MinimalModel6}.}

{It is well-known that the prototypical two- and three-site Kitaev chain, realized
through spin-polarized semiconductor QDs coupled via an \textit{s}-wave
superconducting bridge, supports isolated PMM states at well-separated
QDs with grounded QD configurations and balanced quantum processes
of electron cotunneling (ECT) and crossed Andreev reflection (CAR)
\cite{Flensberg_2012(Poor),Kouwenhoven2023,Jelena1,Jelena2,Poor10}.} Particularly at this
\textit{sweet spot} parameter regime, PMMs emerge
as spatially distributed zero-energy modes with non-local correlations
across QDs. However, their non-topological nature manifests through
sensitivity to local electrostatic perturbations, hybridization-induced
delocalization under gate potential variations and absence of bulk-boundary
correspondence.

{Counterintuitively, despite the lack of true topological protection
in PMM systems, PMMs paradoxically enable enhanced quantum control
capabilities such as direct state manipulation without topological
protection breaking protocols, native initialization and readout of
qubit states, encoding using four PMM modes and demonstration of non-Abelian
statistics through braiding and fusion operations~\cite{Poor3,Poor14,Poor15,Poor16,Poor17,Poor19}.}

{Remarkably, the qubit information is encoded in the fermionic parity
states (even/odd occupation) of Dirac fermions formed by PMM pairs~\cite{Flensberg_2012(Poor),FermionParity,PMMTetron},
and the operational fidelity is ruled by the Majorana quality factor
(also called Majorana Polarization), which is a quantitative metric
characterizing the degree of wavefunction non-locality across the
QDs~\cite{Poor4,Poor13,MinimalModel7,Poor3}.} Noteworthy, such an
approach on the Majorana non-locality feature resembles the situation
of QDs coupled to superconducting wires \cite{Nonlocality_Majorana}.

Given the previously outlined scenario concerning the low stability
and inherent lack of topological protection in PMMs for realizing
fault-tolerant quantum computation, below we focus on a concise review
of the minimal Kitaev chain architecture wherein PMMs manifest as
MBSs localized on distinct and spatially separated QDs. Furthermore,
to delineate the methodology proposed by some of us in Ref. \cite{Sanches_2025}
for leveraging this inherent vulnerability for the Spin-Exchange Induced
Spillover Effect, we develop a comprehensive theoretical framework
based on the Green's function formalism. This framework is employed
to quantitatively evaluate the phenomenon of Spin-Exchange Spillover
on PMMs \cite{Sanches_2025}.

\begin{figure}[!]
\centering\includegraphics[width=1\columnwidth]{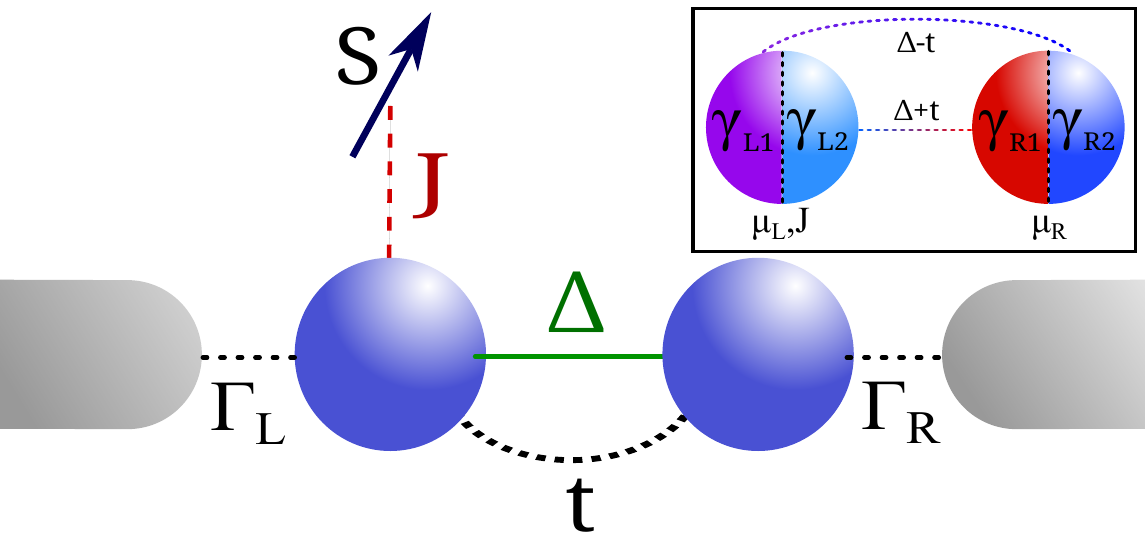} \caption{\label{Fig:Fig.1} {The minimal Kitaev chain made by two quantum
dots (QDs) exhibit electron cotunneling ($t$) and crossed Andreev
reflection ($\Delta$), where at the condition
$t=\Delta$ and chemical potentials $\mu_{L}=\mu_{R}=0$ (inset),
the system functions as a spin-statistics probe for the spin $S$
quasiparticles via the \textquotedblleft Poor Man's Majorana\textquotedblright{}
(PMM) spillover effect. This phenomenon arises from exchange coupling
($J$) fluctuations, mediating a projection of the left PMM onto the
right QD. The spin-statistics signature manifests as a multi-level
spectrum with multiplicity $2S+2$ ($2S+1$) for bosonic (fermionic)
spin $S$ states, localized predominantly on the left dot. Given the
hybridization of both dots with distinct fermionic reservoirs (gray
metallic leads), the asymmetric coupling $\Gamma_{\alpha}=\delta_{\alpha L}\Gamma+\eta$
($\alpha=L(R)$ for left (right) dot), which is considered to mimic
the experimental conditions of Ref.\cite{Kouwenhoven2023}, emerges
as critical in stabilizing the PMMs $\gamma_{L1}$ and $\gamma_{R2}$
under constrained $J$ variations. Consequently, $\gamma_{L1},\gamma_{L2},$
and $\gamma_{R1}$ form an unconventional trimer configuration, where
the zero-energy mode is exclusively hosted by $\gamma_{L1}$ and absent
in $\gamma_{R1}$. This dichotomy stems from the interplay between
$\Gamma$ and $\eta$, which isolates $\gamma_{L1}$ and $\gamma_{R2}.$}}
\end{figure}

The fundamental mechanism for this spillover is the projection of
the PMM wavefunction from its host QD onto the adjacent site. Crucially,
we distinguish this electrostatic mechanism from the primary focus
of this study: spillover induced by a non-electrostatic local potential
shift arising from exchange coupling to a proximal quantum spin, as
proposed in Ref. \cite{Sanches_2025}. Interestingly enough, the system's
interaction with a multi-terminal environment is explicitly incorporated
into our analysis to match the experimental arrangement of Ref.\cite{Kouwenhoven2023}
{[}Fig.\ref{Fig:Fig.1}{]}.

We highlight that the Spin-Exchange Induced Spillover consists in
perturbing locally a QD with a PMM by an exchange coupling $J$ to
induce the PMM wavefunction projection onto the neighboring QD. As
aftermath, such a spillover gives rise to a multi-level structure
spanning $2S+1$ $(2S+2)$ subgap states when a half-integer (integer)
quantum spin $S$ for fermions (bosons) is placed next to the QD hosting
the PMM \cite{Sanches_2025}.

Noteworthy, analogous spin spectroscopies of subgap states based on
spin-polarized QDs or magnetic fields were already performed by other
groups \cite{SpinSpectro1,SpinSpectro2}. Particularly in Ref. \cite{SpinSpectro1},
the technique was applied to superconducting hybrid nanowires, with
the goal of distinguishing ABSs with spin-$0$ from those carrying
spin-$\frac{1}{2}.$

In this work, the setup of Fig.\ref{Fig:Fig.1} already begins with
defined $p$-wave superconductivity (the minimal Kitaev chain) and
if an exchange coupling to an extra quantum spin $S$ is assumed with
a single QD, the PMM spillover induced by $J$ can be applied to unveil
the value of this arbitrary $S,$ in particular via local spectroscopic
detection of subgap states at the QD nearby $S.$ Thus, solely one
extra spin $S$ is enough for local spectral measurements via differential
conductance {[}see details in Sec. 3.4{]}.

Distinctly from our earlier findings \cite{Sanches_2025}, where the
Spin-Exchange Induced Spillover was introduced, in this Topical Review
paper, we also explore conceptually the interplay between the couplings
of the Kitaev dimer to asymmetric metallic reservoirs and an exchange
parameter due to a quantum spin, thus leading to the \textquotedblleft environmentally
induced protection\textquotedblright . This study reveals that PMMs
retain a degree of partial dissipative spectral stabilization against
moderate fluctuations in the exchange coupling strength. Noteworthy,
as we shall see, such a ``protection'' does not correspond to the
``highly desired topological protection'' for the Majorana based
parity-qubit, which is expected to occur in experimental long QD arrays
\cite{MinimalModel3}.

However, before starting into the grounds of the Spin-Exchange Induced
Spillover mechanisms, in the following we summarize the main findings
covering the experimental platforms, wherein the PMM realization is
considered achievable.

\section{Experimental QD Arrays}

In the work by Ref.\cite{Kouwenhoven2023}, Tom Dvir \textit{et al.}
report the experimental realization of a minimal Kitaev chain. This
quantum architecture was fabricated from an InSb nanowire hosting
two gate-defined QDs that are coherently coupled through a superconducting-semiconducting
hybrid segment. This platform successfully mitigates several longstanding
experimental issues, as it enables strong inter-dot hybridization
mediated by CAR, establishes simultaneous coupling between two isolated
spins via both ECT and CAR processes, and permits continuous, \textit{in
situ} electrostatic control over the respective coupling strengths.
This high degree of tunability is facilitated by the discrete, gate-configurable
energy spectra of the two QDs and the mediating ABSs, which function
as independently addressable quantum systems. The coherent hybridization
of these discrete states gives rise to a delocalized, nonlocal fermionic
mode, which supports Majorana-like excitations when electrostatically
tuned to a \textit{sweet spot} regime. Spectroscopic measurements
at this operational regime reveal zero-bias conductance peaks that
demonstrate key signatures consistent with emergent Majorana physics.
This modular approach to constructing Kitaev chains is able to achieve
a high degree of concordance between theoretical predictions and empirical
results, primarily circumventing the complications given by disorder,
which plagues nanowire implementations. Scaling this platform to extended
one-dimensional arrays presents a promising pathway for the future
realization of topologically protected Majorana zero modes.

In a complementary study, in Ref.\cite{MinimalModel2} Sebastiaan
L. D. ten Haaf \textit{et al.} report the implementation of a two-site
Kitaev chain via the coherent coupling of two spin-polarized QDs engineered
within an InSb/As two-dimensional electron gas (2DEG). The authors
employed precise electrostatic gating to tune the hybridized system
into designated \textit{sweet spot} regime, where correlated zero-bias
conductance peaks were spectroscopically resolved. Beyond standard
electrostatic control, the planar 2DEG geometry was leveraged to demonstrate
that access to the \textit{sweet spot} can also be achieved through
in-plane rotations of the magnetic field vector. While the formation
of localized MBSs formally requires a substantial Zeeman energy, it
was counterintuitively observed that key signatures (most notably
strongly correlated zero-bias peaks) persist upon reduction of the
magnetic field to zero. The evolution of the many-body energy spectrum
with applied magnetic field provides complementary spectroscopic data,
which was analyzed to quantify the degree of hybridization between
emergent Majorana states using the metric of Majorana Polarization
\cite{Poor4,Poor13,Poor3}.

Additionally, the implementation of fermionic parity readout within
a minimal Kitaev chain architecture is needed for quantum computing
purposes and demands advanced experimental techniques. Particularly,
the integration of complementary measurement methods, such as charge
sensing, parity polarization, and transport, can be employed to guide
precise electrostatic tuning of the Kitaev chain as demonstrated in
Ref.\cite{MinimalModel5} by Nick van Loo \textit{et al.} These diagnostic
tools provide consistent indicators of operational proximity to the
\textit{sweet spot}. A critical advantage of the capacitive readout
method proposed is its robustness to precise parameter tuning, i.e.,
the quantum capacitance retains its sensitivity to the fermionic parity
eigenvalue even under detuning from exact energy degeneracy. The described
parity-to-charge conversion technique is specifically engineered for
integration into scalable qubit architectures. In a proposed device
geometry featuring two Kitaev chains coupled via a common superconducting
segment, each chain exhibits a distinct parity-dependent quantum capacitance.
This property enables single-shot discrimination across the complete
four-state parity basis, thereby facilitating the detection of the
system leakage out. This methodology addresses a fundamental challenge
in the development of Majorana-based quantum information processing
by enabling rapid and high-fidelity parity readout. The simultaneous
deployment of a high-sensitivity charge sensor enables the experimental
confirmation of local charge neutrality concomitant with parity eigenstates,
a defining spectroscopic hallmark of MBSs. The establishment of this
metrological capability renders mature gate-based control methodologies,
originally developed within the spin qubit paradigm, directly transferable
to this platform.

In Ref.\cite{MinimalModel7}, by Alberto Bordin \textit{et al.}, the
experimental realization of both two- and three-site Kitaev chains
was demonstrated within a nanowire architecture. In this work, the
authors reveal full control over the superconducting phase difference
($\varphi$). Particularly, such a parameter was modulated through
application of an out-of-plane magnetic field. Notably, electrostatic
gating techniques enabled access to configurations with $\varphi\approx0$
or $\varphi\approx\pi$ even at zero applied magnetic field. This
capability establishes a viable pathway for phase control in extended
Kitaev chains through selective\textit{ sweet spot} operation, obviating
the need for intricate flux-bias control circuitry. At $\varphi=0$,
a detailed spectroscopic analysis of the three-site chain was conducted
under systematically applied local and global electrostatic perturbations.
The measured energy spectra demonstrate exceptional agreement with
the theoretical Kitaev Hamiltonian. The robustness of the emergent
MBSs localized at the chain edges was further probed against external
perturbations via coupling to an additional QD serving as a spectroscopic
probe. This test protocol provides heightened sensitivity to the wavefunction
overlap between the two unpaired Majoranas on a single device side,
even when conventional tunneling spectroscopy remains insensitive
to such hybridization. Although wavefunction overlap could theoretically
induce splitting of the zero-bias peak, no spectral splitting was
observed under optimal \textit{sweet spot} conditions.

Regarding the experimental realization in Ref.\cite{MinimalModel3},
by Alberto Bordin \textit{et al.}, a strongly hybridized three-QD
array was engineered via coherent inter-dot coupling mediated by both
CAR and ECT processes. The device architecture is further configured
to concurrently host two adjacent and minimal two-site Kitaev chains.
Crucially, when the three QDs are electrostatically tuned into mutual
energy resonance, the system manifests a spectroscopic signature consistent
with the theoretical spectrum of a three-site Kitaev chain. This configuration
establishes a versatile platform for a systematic investigation of
MBS stability within an extended chain, particularly against local
fluctuations in the chemical potential and against variations in the
relative strengths of the CAR and ECT coupling amplitudes. This advancement
directly addresses a fundamental limitation of minimal (two-site)
implementations, where the finite wavefunction overlap of the MBSs
constitute the main origin of decoherence. Consequently, scaling the
system to a three-site chain significantly enhances the robustness
of the subsystem against disorder in local potentials.

In a related investigation by Sebastiaan L. D. ten Haaf \textit{et
al.} \cite{MinimalModel4} of a hybridized three-QD array, implemented
within an InSb/As 2DEG, a comprehensive spectroscopic analysis of
both bulk and edge states is presented. A defining characteristic
of this architecture is the integration of individual ohmic contacts
to each QD, which facilitates direct local tunneling spectroscopy
and enables a site-resolved quantification of the local density of
states. Through independent electrostatic control of the interdot
coupling strengths and the superconducting phase difference, the system
was tuned into a regime exhibiting zero-bias conductance peaks localized
on the terminal QDs. These correlated edge zero-bias-peaks coincide
with a significant suppression of subgap states, manifesting as a
hard superconducting gap within the spectral signature of the central
QD. The magnitude of this bulk spectral gap is tunable via the phase
difference, providing a direct empirical correlation between the spectral
isolation of the zero-energy edge modes and their enhanced robustness
to local electrostatic perturbations. The spectroscopic access to
all sites permits a direct spatial mapping of the Majorana wavefunction
probability distribution across the chain, allowing its evolution
to be tracked as a function of applied gate voltages. Furthermore,
an experimental phase diagram was constructed for this finite-length
Kitaev chain, identifying regions in parameter space defined by the
coupling amplitudes and on-site potentials where zero-bias peaks become
stabilized. As aftermath, this analysis demonstrates that the stability
region expands upon scaling from a two-site to a three-site chain.

\section{Theoretical QD Arrays}

The theoretical framework underpinning the experiments reported in
Refs. \cite{Kouwenhoven2023,MinimalModel2,MinimalModel5,MinimalModel7,MinimalModel3,MinimalModel4}
is a model that explicitly incorporates spin degrees of freedom and
Coulomb interactions within the QDs, while permitting strong hybridization
with a proximate superconductor \cite{Poor3}. The corresponding system
Hamiltonian is given by
\begin{equation}
\mathcal{H}_{0}=\mathcal{H}_{\text{QDs}}+\mathcal{H}_{\text{ABS}}+\mathcal{H}_{T},\label{eq:TotalH}
\end{equation}
where the Hamiltonian for the QDs is
\begin{equation}
\mathcal{H}_{\text{QDs}}=\sum_{\sigma,\alpha}\varepsilon_{\alpha\sigma}n_{\alpha\sigma}+\sum_{\alpha}U_{\alpha}n_{\alpha\uparrow}n_{\alpha\downarrow},\label{eq:Hubbard}
\end{equation}
where $d_{\alpha\sigma}^{\dagger}$ creates an electron with spin
$\sigma=\uparrow,\downarrow$ on QD $\alpha=L,R$, and $n_{\alpha\sigma}=d_{\alpha\sigma}^{\dagger}d_{\alpha\sigma}$
is the corresponding number operator. The single-particle energy $\varepsilon_{\alpha\sigma}=\varepsilon_{\alpha}\pm\Delta_{Z}^{\alpha}/2$
includes the orbital energy $\varepsilon_{\alpha}$ (the chemical
potential) and the Zeeman splitting $\Delta_{Z}^{\alpha}.$ The term
$U_{\alpha}$ denotes the on-site Coulomb charging energy.

The coupling between the QDs is mediated by a discrete ABS within
the superconducting segment. This ABS may arise from a proximitized
semiconductor nanowire. Its Hamiltonian is
\begin{equation}
\mathcal{H}_{\text{ABS}}=\sum_{\sigma}\varepsilon_{\text{ABS},\sigma}n_{\text{ABS},\sigma}+(\tilde{\Delta}c_{\uparrow}^{\dagger}c_{\downarrow}^{\dagger}+\text{{H.c.}}),\label{eq:H_ABS}
\end{equation}
where $c_{\sigma}^{\dagger}$ is the creation operator for the ABS,
$n_{\text{ABS},\sigma}=c_{\sigma}^{\dagger}c_{\sigma}$, and $\varepsilon_{\text{ABS},\sigma}=\varepsilon_{\text{ABS}}\pm\Delta_{Z}^{\text{ABS}}/2$.
Coulomb interactions within the ABS are neglected, justified by its
strong capacitive coupling to a grounded bulk superconductor that
provides efficient charge screening and $\tilde{\Delta}$ is the s-wave
superconducting pairing.

Tunneling between the QDs and the ABS is described by the Hamiltonian
\begin{align}
\mathcal{H}_{T} & =\sum_{\sigma}s_{\sigma}(t_{\text{SO},L}d_{L\sigma}^{\dagger}c_{\bar{\sigma}}+t_{\text{SO},R}c_{\sigma}^{\dagger}d_{R\bar{\sigma}})+\sum_{\sigma,\alpha}t_{\alpha}d_{\alpha\sigma}^{\dagger}c_{\sigma}\nonumber \\
 & +\text{H.c.}.\label{eq:Tunneling}
\end{align}
Here, $s_{\uparrow,\downarrow}=\pm1$. The amplitude $t_{\alpha}$
denotes spin-conserving tunneling between QD $\alpha$ and the ABS,
while $t_{\text{SO},\alpha}$ represents the spin-orbit-induced spin-flip
tunneling amplitude. This term originates from a spin-orbit field
oriented along the $y$-axis, orthogonal to the external Zeeman field.

\subsection{PMM Quantum Computation }

The quantum information encoded in a pair of MBSs is topologically
protected due to its association with the fermion parity operator,
a global symmetry that is conserved in topological superconducting
systems. Consequently, the minimal architecture for a functional qubit
necessitates a two-level subspace defined by the degenerate states
of four MBSs \cite{RoadmapTetron2025}, wherein the total fermion
parity is constrained to a fixed eigenvalue \cite{Flensberg-2012,Alicea-2012}.
Arbitrary single-qubit rotations, which are not topologically protected
operations, can be executed by controllably lifting the degeneracy
through the introduction of an energy splitting associated with the
occupation of the non-local fermionic mode defined by specific MBS
pairs. This tunable coupling between MBSs can be engineered via two
primary physical mechanisms: direct hybridization of their wavefunctions
\cite{Hell1} or through Coulomb-mediated interactions \cite{Hell2}.
Thereby, on the grounds of PMM-based qubits, a coherent coupling between
two distinct PMM systems was proposed in Ref.\cite{Poor3}, where
the coupling can be performed by direct tunnel coupling between a
QD in each system or an indirect coupling mediated by a shared superconducting
reservoir. In the aforementioned paper, Athanasios Tsintzis \textit{et al.}
discuss several aspects covering quantum computing protocols with
PMMs, such as non-abelian operations, braiding, fusion, initialization
and readout of PMM-based qubits. Therefore, we summarize below the
main results concerning such protocols.

The defining characteristic of non-Abelian anyons is that their exchange
statistics are governed by non-commutative representations of the
braid group \cite{DasSarma-2008}, rendering the outcome of sequential
exchange operations order-dependent. The topological robustness of
these braiding operations, wherein, for an ideal system, the result
depends solely on the topological class of the braid and not on its
specific geometric realization, constitutes the fundamental principle
underpinning topological quantum computation \cite{DasSarma-2008,Nayak-2015,Marra-2022}.
Experimental verification of the non-Abelian statistics of MBSs could,
in principle, be achieved through their physical translocation, with
recent proposals outlining such protocols in QD-based Kitaev chains
\cite{Boross}. Notably, Athanasios Tsintzis \textit{et al.} performed
an analysis focused on quantifying the deviation from the theoretical
braiding transformations induced by non-ideal MBS characteristics,
as quantified by the Majorana Polarization. They found a critical
distinction from fusion rules: while the implemented operation remains
non-Abelian, its unitary transformation converges asymptotically towards
the topologically protected result solely in the limit of a unit Majorana
Polarization. Consequently, the authors emphasize that braiding experiments
constitute the definitive benchmark for assessing the topological
equivalence between PMM systems and genuine topologically ordered
MBSs.

The protocol of spatially coalescing non-Abelian anyons to perform
a projective measurement of their collective state is termed fusion,
with the permissible outcomes dictated by their characteristic fusion
rules \cite{DasSarma-2008}, once MBSs constitute a physical realization
of Ising anyon statistics \cite{DasSarma-2008}. Their fusion rules
dictate that a pair of MBSs may either fuse to the vacuum channel
(corresponding to a measurement outcome of even fermion parity, or
an unoccupied non-local fermionic state in the aforementioned formalism)
or to the fermion channel (corresponding to a measurement outcome
of odd fermion parity or an occupied fermionic state). While the fusion
protocol for PMMs presents a demonstrably feasible experimental pathway,
it exhibits a critical insensitivity to the degree of Majorana character,
as quantified by the Majorana Polarization. The authors analysis demonstrates
that the fusion outcome is invariant with respect to the Majorana
Polarization, contingent upon the experimental capability to fine-tune
the system to a parameter regime where the uncoupled PMM subsystems
exhibit a precise even-odd fermion parity degeneracy.

Interestingly enough, Athanasios Tsintzis \textit{et al.} establish
that the inherent non-topological nature of PMMs, while precluding
full topological protection, simultaneously presents a functional
advantage by enabling simplified qubit initialization and readout
protocols. The authors present methodologies for quantifying the state
lifetime of these excitations, a parameter that fundamentally limits
the coherence time and defines the relaxation rate of a PMM-based
qubit. The execution of these experimental protocols necessitates
the capacitive coupling of, at least, one QD within the PMM architecture
to a high-resolution charge sensor, such as a quantum point contact
or a single-electron transistor or its integration into a circuit
configured for quantum capacitance measurements.

\subsection{The Extra Spin and Multi-Terminal Couplings}

Motivated by recent experimental work reported in Refs.~\cite{Kouwenhoven2023,MinimalModel2},
we revisit the system schematically illustrated in Fig.~\ref{Fig:Fig.1},
which was proposed by some of us in Ref. \cite{Sanches_2025} in treating
the minimal Kitaev chain in the presence of an exchange coupling.
A key feature of our model is the inclusion of multi-energy level
effects arising from a quantum spin $S$. The theoretical description
of this interaction accounts for the extra spin $S$ coupled to the
magnetic field $\vec{B}=B\hat{z},$ which reads

\begin{equation}
\mathcal{H}_{\text{\text{{S}}}}=-\gamma_{S}\vec{S}.\vec{B}=-\gamma_{S}BS^{z}=-\Delta_{Z}^{S}\sum_{m}m|m\rangle\langle m|\label{eq:HS}
\end{equation}
expressed in terms of $\gamma_{S}$ and $\Delta_{Z}^{S}=\gamma_{S}B$
standing for the gyromagnetic ratio and Zeeman splitting for the spin
$S,$ respectively, where we used the standard expansion

\begin{equation}
S^{z}=\sum_{m}m|m\rangle\langle m|,\label{eq:SzComponents}
\end{equation}
with $m\in\{-S,-S+1,\dots,S-1,S\}.$ Later on, we couple the quantum
spin $S$ to the left QD spin $s=\frac{1}{2},$ which is performed
via the anisotropic spin-spin coupling Hamiltonian obeying the Heisenberg-like
model

\begin{equation}
\mathcal{H}_{\text{\text{{Ss}}}}=\mathcal{H}_{\text{\text{{Ising}}}}+\frac{J_{\bot}}{2}(S^{+}s_{L}^{-}+S^{-}s_{L}^{+}),\label{eq:HSs}
\end{equation}
wherein the Ising term
\begin{equation}
\mathcal{H}_{\text{\text{{Ising}}}}=JS^{z}s_{L}^{z}\label{eq:IsingH}
\end{equation}
describes the coupling between the spin-$z$ operators via the exchange
parameter $J,$ with similar expansion
\begin{equation}
s_{L}^{z}=\frac{1}{2}\sum_{\sigma}\sigma d_{L\sigma}^{\dagger}d_{L\sigma}\label{eq:RQDspin}
\end{equation}
for the left QD, while the anisotropic exchange coupling $J_{\bot}$
rules the spin-flip processes between $S$ and the left QD, being
$S^{\pm}$ and $s_{L}^{\pm}$ the ladder operators for the quantum spin
$S$ and left QD, respectively.

We are specially interested in the system spinless regime, i.e., when
the constraints $|\varepsilon_{\alpha\uparrow}|,|t_{\alpha}|,|t_{\text{SO,}\alpha}|,|\Delta_{Z}^{\text{ABS}}|,|\Delta_{Z}^{S}|\ll|\tilde{\Delta}|$
and $|\Delta_{Z}^{L,R}|$ (large Zeeman field over the QDs) are fulfilled.
Within such a situation, the occupation of the excited spin-down channel
($\sigma=\downarrow$) can be safely neglected at the QDs ($<d_{\alpha\downarrow}^{\dagger}d_{\alpha\downarrow}>=0$),
as well as the spin-flip processes between the quantum spin $S$ and
the left QD ($J_{\bot}=0$), once the applied strong magnetic field
spin polarizes the electrons.

Particularly for $J_{\bot}=J,$ Eq.(\ref{eq:HSs}) recovers the Heisenberg
Hamiltonian, but it corresponds to the opposite limiting case (weak Zeeman field over the QDs) here
considered and does not belong to
the current analysis. Consequently, the full interacting model of
Eq.(\ref{eq:TotalH}) reduces to an effective low-energy description
that is both spinless and noninteracting Hamiltonian, which hosts,
as we shall see, PMM states in the so-called \textit{sweet spot} \cite{Flensberg_2012(Poor),Sanches_2025}.

In summary, the system then operates within a strong Zeeman splitting
condition at the QDs, which restricts transport to the spin-up channel
($\sigma=\uparrow$). Thus, the effective coupling between QDs $L$
and $R$, mediated by the ABS {[}Eq.(\ref{eq:H_ABS}){]}, can be treated
within a second-order perturbation framework \cite{Poor3}. As a result,
this two-QD configuration forms the minimal Kitaev chain, with the
dots coupled via ECT, characterized by the hopping amplitude $t$,
and CAR, mediated by the superconducting pairing term $\Delta$ \cite{Flensberg_2012(Poor)},
whose original Hamiltonian $\mathcal{H}_{\text{0}}+\mathcal{H}_{\text{\text{{S}}}}+\mathcal{H}_{\text{\text{{Ss}}}}$
in the presence of the exchange coupling $J$ can be mapped onto
\begin{eqnarray}
{\cal {H}} & = & (\mu_{L\uparrow}+\frac{J}{2}S^{z})d_{L\uparrow}^{\dagger}d_{L\uparrow}+\mu_{R\uparrow}d_{R\uparrow}^{\dagger}d_{R\uparrow}\nonumber \\
 & + & (td_{L\uparrow}d_{R\uparrow}^{\dagger}+\Delta d_{L\uparrow}d_{R\uparrow}+\text{{H.c.}}),\label{eq:H}
\end{eqnarray}
where $\mu_{L\uparrow(R\uparrow)}$ represents the renormalized chemical
potential \cite{Poor3} for the QD $\alpha=L,R$ \cite{Sanches_2025}.
Thereby, based on Eq.(\ref{eq:H}), we shall see the constraints to
be fulfilled by the model to host PMMs in the absence or presence
of the exchange coupling, being the latter the case determinant
to the Spin-Exchange Induced Spillover emergence.{{} }

Additionally, the electronic operators of the chain can be expressed
in the\textit{ }MBS basis $\gamma_{L1(L2)}$ and $\gamma_{R1(R2)}$
for the left and the right QDs, respectively. This change of basis
is performed according to the following relations
\begin{equation}
d_{L\uparrow}=\frac{1}{\sqrt{2}}(\gamma_{L1}+i\gamma_{L2})\label{eq:dL}
\end{equation}
and
\begin{equation}
d_{R\uparrow}=\frac{1}{\sqrt{2}}(\gamma_{R1}+i\gamma_{R2}),\label{eq:dR}
\end{equation}
which yield
\begin{align}
{\cal {H}} & =(\mu_{L\uparrow}+\frac{J}{2}S^{z})(\frac{1}{2}+i\gamma_{L1}\gamma_{L2})+\mu_{R\uparrow}(\frac{1}{2}+i\gamma_{R1}\gamma_{R2})\nonumber \\
 & +i(\Delta-t)\gamma_{L1}\gamma_{R2}+i(\Delta+t)\gamma_{L2}\gamma_{R1}.\label{eq:MajoranaChain}
\end{align}
The \textit{``Majorana chain regime''} ($t=\Delta$) is characterized
by a tight-binding Hamiltonian constructed from Majorana operators,
such as $\gamma_{L1}$, $\gamma_{L2}$, and $\gamma_{R1}$ as the
trimer configuration, where $\mu_{L\uparrow}$ or (and) $J$ should
be finite (finites) and $\mu_{R\uparrow}=0$ given by
\begin{equation}
{\cal {H}}_{\text{{trimer}}}=(\mu_{L\uparrow}+\frac{J}{2}S^{z})(\frac{1}{2}+i\gamma_{L1}\gamma_{L2})+i2\Delta\gamma_{L2}\gamma_{R1},\label{eq:Trimer}
\end{equation}
as well as with $\gamma_{L2}$ and $\gamma_{R1}$ being the dimer
configuration. The dimer case corresponds to the \textit{sweet spot}
condition~\cite{Flensberg_2012(Poor)}, defined by $\mu_{\alpha\uparrow}=\mu_{\bar{\alpha}\uparrow}=J=0$
(with $\bar{\alpha}=R(L)$ for $\alpha=L(R)$) together with $t=\Delta,$
namely
\begin{equation}
{\cal {H}}_{\text{{dimer}}}=i2\Delta\gamma_{L2}\gamma_{R1},\label{eq:Dimer}
\end{equation}
that also supports spatially localized PMM zero-modes $\gamma_{L1}$
and $\gamma_{R2}$ at the left and right QDs, respectively, as demonstrated
in Refs.~\cite{Flensberg_2012(Poor),Sanches_2025,MinimalModel2,Kouwenhoven2023}.

The exchange coupling $J$ plays a dual role as both an effective
chemical potential for the left QD and a driver of PMM spillover effect.
This spin-exchange hybridization mechanism produces distinct spectroscopic
signatures depending on the quantum statistics of the spin $S$:
\begin{itemize}
\item fermionic statistics yields $2S+1$ subgap states;
\item bosonic statistics yields $2S+2$ subgap states.
\end{itemize}
The resulting interference between exchange energy $J$ and superconducting
pairing $\Delta$ enables simultaneous control of PMM spatial distribution
and probing of fundamental spin properties, complementing conventional
gate-voltage tuning approaches. The spin-dependent spillover phenomenon
provides a powerful tool for investigating Majorana hybridization
dynamics while revealing essential information about the underlying
spin degrees of freedom through characteristic spectral features in
the subgap regime.

In contrast to the eigenvalue-based theoretical approach developed
in Ref.~\cite{Flensberg_2012(Poor)}, where PMM spillover is characterized
through wavefunction projection under local perturbations in the absence
of a multi-terminal coupling, we adopt here the Green's functions
(GFs) formalism \cite{Sanches_2025,Zubarev-1960,Flensberg-book}.
This framework enables a comprehensive description of the spillover
phenomenon through analysis of the retarded GFs $\langle\langle A_{\alpha};B_{\alpha}\rangle\rangle$
for the QD $\alpha$, in terms of the the energy variable $\omega.$

The complete characterization of PMM spillover dynamics requires evaluation
of both normal and anomalous spectral functions ${\cal {A}}_{A_{\alpha}B_{\alpha}}(\omega)$.
We begin with the normal GFs as follows
\begin{equation}
{\cal {A}}_{d_{\alpha\uparrow}d_{\alpha\uparrow}^{\dagger}}(\omega)=-\frac{1}{\pi}\text{Im}\langle\langle d_{\alpha\uparrow};d_{\alpha\uparrow}^{\dagger}\rangle\rangle\label{eq:DOSddx}
\end{equation}
and
\begin{equation}
{\cal {A}}_{d_{\alpha\uparrow}^{\dagger}d_{\alpha\uparrow}}(\omega)=-\frac{1}{\pi}\text{Im}\langle\langle d_{\alpha\uparrow}^{\dagger};d_{\alpha\uparrow}\rangle\rangle.\label{eq:DOSdxd}
\end{equation}

Additionally, we must include the anomalous GFs
\begin{equation}
{\cal {A}}_{d_{\alpha\uparrow}^{\dagger}d_{\alpha\uparrow}^{\dagger}}(\omega)=-\frac{1}{\pi}\text{Im}\langle\langle d_{\alpha\uparrow}^{\dagger};d_{\alpha\uparrow}^{\dagger}\rangle\rangle\label{eq:DOSdxdx}
\end{equation}
and
\begin{equation}
{\cal {A}}_{d_{\alpha\uparrow}d_{\alpha\uparrow}}(\omega)=-\frac{1}{\pi}\text{Im}\langle\langle d_{\alpha\uparrow};d_{\alpha\uparrow}\rangle\rangle.\label{eq:DOSdd}
\end{equation}
These quantities are essential for determining the MBS component
\begin{equation}
{\cal {A}}_{\gamma_{\alpha j}}(\omega)=-\frac{1}{\pi}\text{Im}\langle\langle\gamma_{\alpha j};\gamma_{\alpha j}\rangle\rangle\label{eq:DOSMF}
\end{equation}
associated with each QD. For the Majorana operators $\gamma_{L1(L2)}$
and $\gamma_{R1(R2)}$, the GF can be expressed as
\begin{eqnarray}
\langle\langle\gamma_{\alpha j};\gamma_{\alpha j}\rangle\rangle & = & \frac{1}{2}[\langle\langle d_{\alpha\uparrow};d_{\alpha\uparrow}^{\dagger}\rangle\rangle+\langle\langle d_{\alpha\uparrow}^{\dagger};d_{\alpha\uparrow}\rangle\rangle\nonumber \\
 & + & \epsilon_{j}(\langle\langle d_{\alpha\uparrow}^{\dagger};d_{\alpha\uparrow}^{\dagger}\rangle\rangle+\langle\langle d_{\alpha\uparrow};d_{\alpha\uparrow}\rangle\rangle)],\label{eq:Expansion}
\end{eqnarray}
where $\epsilon_{j}=+1,-1$ for $j=1,2$.

The right-hand side GFs in Eq. (\ref{eq:Expansion}) can be computed
using the standard equation-of-motion (EOM) method \cite{Zubarev-1960,Flensberg-book},
whose implementation is summarized in the following relation
\begin{eqnarray}
(\omega+i\Gamma_{\alpha})\langle\langle A_{\alpha};B_{\alpha}\rangle\rangle=\langle[A_{\alpha},B_{\alpha}]_{+}\rangle+\langle\langle[A_{\alpha},{\cal {\cal {H}}}];B_{\alpha}\rangle\rangle,\nonumber \\
\label{eq:EOM}
\end{eqnarray}
where $\Gamma_{\alpha}=\delta_{\alpha L}\Gamma+\eta$ ($\alpha=L(R)$
for left (right) QD) mimics the broadening arising from the multi-terminal
environment as in Ref.\cite{Kouwenhoven2023} and $\delta_{\alpha L}$
denotes the Kronecker delta. Noteworthy, analogously to our theoretical
framework, the non-Hermitian treatment of the two-site Kitaev chain
discussed in Ref. \cite{NonHermitian1} also accounts for an external
environment. However, we distinctly preserve the Hermiticity of ${\cal {H}}$
and insert the outside world effects into the evaluation of the system
GFs instead.

By applying the EOM technique to Eq.(\ref{eq:H}) for $t\neq\Delta$
we find
\begin{eqnarray}
\langle\langle d_{\alpha\uparrow};d_{\alpha\uparrow}^{\dagger}\rangle\rangle & = & \frac{1}{2S+1}\sum_{m}\frac{1}{\omega+i\Gamma_{\alpha}-\mu_{\alpha\uparrow}-\frac{Jm}{2}\delta_{\alpha L}-\Sigma_{\alpha}^{+}},\nonumber \\
\label{eq:GF1}
\end{eqnarray}
\begin{eqnarray}
\langle\langle d_{\alpha\uparrow}^{\dagger};d_{\alpha\uparrow}\rangle\rangle & = & \frac{1}{2S+1}\sum_{m}\frac{1}{\omega+i\Gamma_{\alpha}+\mu_{\alpha\uparrow}+\frac{Jm}{2}\delta_{\alpha L}-\Sigma_{\alpha}^{-}},\nonumber \\
\label{eq:GF2}
\end{eqnarray}
\begin{eqnarray}
\langle\langle d_{\alpha\uparrow}^{\dagger};d_{\alpha\uparrow}^{\dagger}\rangle\rangle & = & \frac{\eta_{\alpha}}{2S+1}\sum_{m}\frac{2t\Delta K_{\alpha}^{-}}{\omega+i\Gamma_{\alpha}+\mu_{\alpha\uparrow}+\frac{Jm}{2}\delta_{\alpha L}-\Sigma_{\alpha}^{-}},\nonumber \\
\label{eq:GF3}
\end{eqnarray}
and
\begin{eqnarray}
\langle\langle d_{\alpha\uparrow};d_{\alpha\uparrow}\rangle\rangle & = & \frac{\eta_{\alpha}}{2S+1}\sum_{m}\frac{2t\Delta K_{\alpha}^{+}}{\omega+i\Gamma_{\alpha}-\mu_{\alpha\uparrow}-\frac{Jm}{2}\delta_{\alpha L}-\Sigma_{\alpha}^{+}},\nonumber \\
\label{eq:GF4}
\end{eqnarray}
where we employed the spectral decomposition $\langle\langle A;B\rangle\rangle=\sum_{m}\langle\langle A\left|m\right\rangle \left\langle m\right|;B\rangle\rangle$,
with the thermal average $\left\langle \left|m\right\rangle \left\langle m\right|\right\rangle =\frac{1}{2S+1}$
and $\eta_{\alpha}=-1,+1$ for $\alpha=L,R$, respectively (see details
in the Appendix). The self-energy correction arising from the couplings
$t$, $\Delta$ and $J$ is given by
\begin{equation}
\Sigma_{\alpha}^{\pm}=\tilde{K}_{\bar{\alpha}}^{\pm}+(2t\Delta)^{2}K_{\bar{\alpha}}K_{\alpha}^{\pm},\label{eq:SelfAlpha}
\end{equation}
with
\begin{eqnarray}
\tilde{K}_{\alpha}^{\pm}=\frac{(\omega+i\Gamma_{\alpha})(t^{2}+\Delta^{2})\pm(\mu_{\alpha\uparrow}+\frac{Jm}{2}\delta_{\alpha L})(t^{2}-\Delta^{2})}{(\omega+i\Gamma_{\alpha})^{2}-(\mu_{\alpha\uparrow}+\frac{Jm}{2}\delta_{\alpha L})^{2}},\nonumber \\
\label{eq:SE1}
\end{eqnarray}
\begin{equation}
K_{\alpha}=\frac{\omega+i\Gamma_{\alpha}}{(\omega+i\Gamma_{\alpha})^{2}-(\mu_{\alpha\uparrow}+\frac{Jm}{2}\delta_{\alpha L})^{2}}\label{eq:SE2}
\end{equation}
and
\begin{equation}
K_{\alpha}^{\pm}=\frac{K_{\bar{\alpha}}}{\omega+i\Gamma_{\alpha}\pm\mu_{\alpha\uparrow}\pm\frac{Jm}{2}\delta_{\alpha L}-\tilde{K}_{\bar{\alpha}}^{\mp}}.\label{eq:SE3}
\end{equation}

The Majorana GF $\langle\langle\gamma_{\alpha j};\gamma_{\alpha j}\rangle\rangle$
in Eq. (\ref{eq:Expansion}) incorporates multiple quantum processes
through its operator expansion, including electron ($\langle\langle d_{\alpha\uparrow};d_{\alpha\uparrow}^{\dagger}\rangle\rangle$)
and hole ($\langle\langle d_{\alpha\uparrow}^{\dagger};d_{\alpha\uparrow}\rangle\rangle$)
propagation terms, along with their corresponding local Andreev reflections
($\langle\langle d_{\alpha\uparrow}^{\dagger};d_{\alpha\uparrow}^{\dagger}\rangle\rangle$
and $\langle\langle d_{\alpha\uparrow};d_{\alpha\uparrow}\rangle\rangle$).
The sign variation of the coefficient $\epsilon_{j}$ with Majorana
index $j$ induces quantum interference between these transport channels,
generating characteristic resonant features in the spectral function
${\cal {A}}_{\gamma_{\alpha j}}(\omega)$.

The interference manifests differently across the configurations of
the \textit{``Majorana chain regime''}: (i) the trimer ($\gamma_{L1},\gamma_{L2},\gamma_{R1}$)
and (ii) the dimer ($\gamma_{L2},\gamma_{R1}$), with the PMMs ($\gamma_{L1},\gamma_{R2}$)
as the \textit{sweet spot}. The resulting spectral features, including
\textit{bonding}, \textit{antibonding} and \textit{nonbonding} molecular
states along with zero-modes all emerge from the quantum interference
processes encoded in ${\cal {A}}_{\gamma_{\alpha j}}(\omega)$.

The Majorana GF formalism thus provides a complete description of
both the fundamental transport processes and their interference-induced
spectral signatures across different Majorana configurations. Below,
we analyze the \textit{``Majorana chain regime}'' with $\Gamma=1$
as the energy scale, $\eta=0.1,$ $t=\Delta=1.5,$ with $\mu_{L\uparrow}=\mu_{L}$
and $\mu_{R\uparrow}=\mu_{R}$ for a sake of simplicity. However,
in some places of our numerical analysis, we switch to realistic units
just to establish a parallel to experimental orders of magnitude in
the system degrees of freedom. To this end, we estimate $t=\Delta\approx12\text{{\ensuremath{\mu}eV}}$
from the experiment of Ref.\cite{Kouwenhoven2023}, thus leading to
$\Gamma=8{\mu eV},$ $\Gamma_{L}=8.8{\mu eV}$ and $\Gamma_{R}=0.8{\mu eV}.$

\begin{figure}[!]
\centering\includegraphics[width=1\columnwidth]{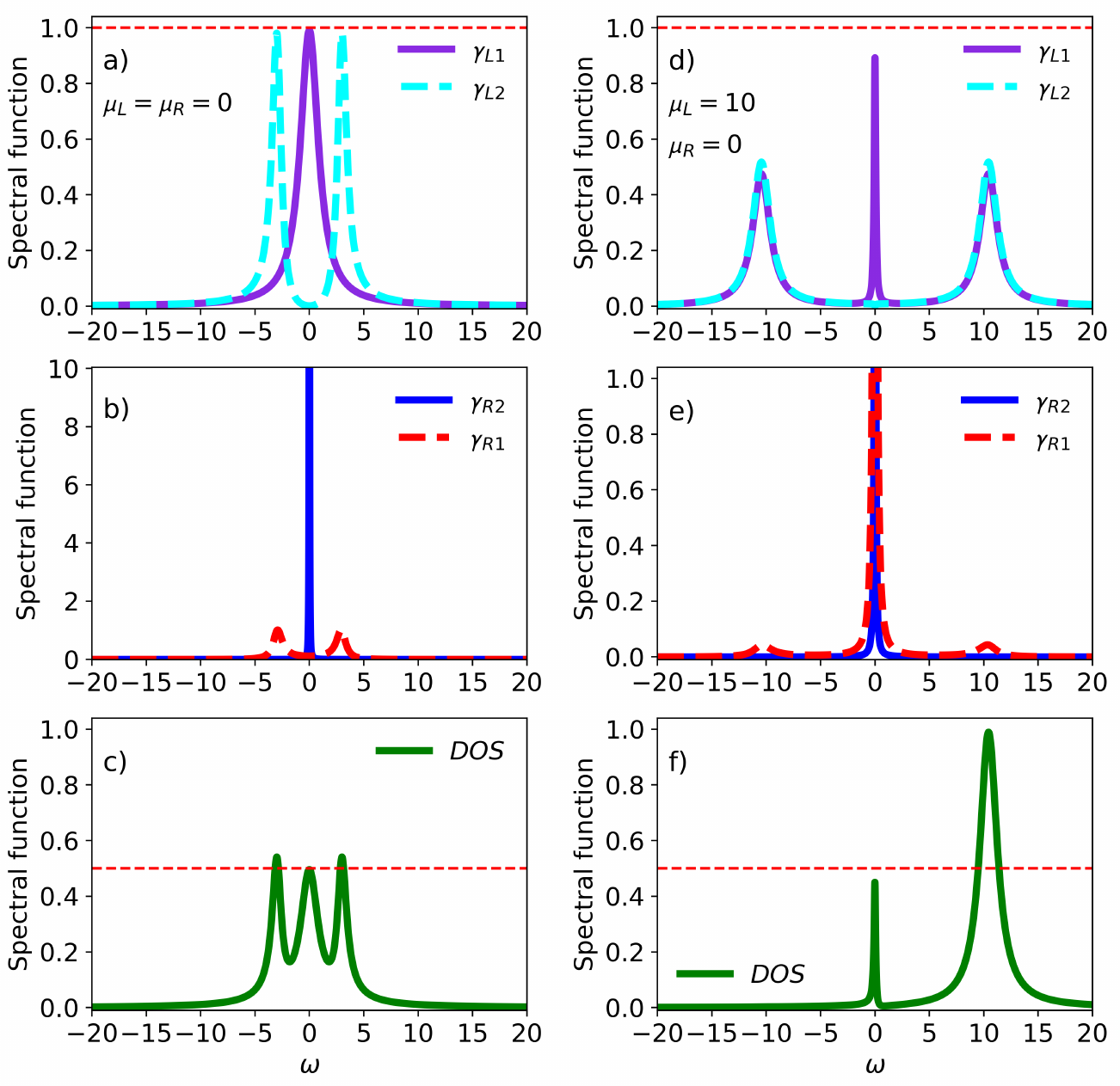} \caption{\label{Fig:Fig.2}{Chemical Potential Induced Spillover: The Majorana
($\pi\Gamma{\cal {A}}_{\gamma_{\alpha j}}$) and electronic ($\pi\Gamma{\cal {A}}_{d_{L\uparrow}d_{L\uparrow}^{\dagger}}$)
spectral functions are analyzed. (a)-(c) At the \textit{sweet spot}
($t=\Delta$ with vanishing local potentials $\mu_{L}=\mu_{R}=J=0$),
a zero-energy mode emerges in $\pi\Gamma{\cal {A}}_{\gamma_{L1}}$
{[}panel (a), where $\pi\Gamma{\cal {A}}_{\gamma_{L1}}(0)=1${]} and
in $\pi\Gamma{\cal {A}}_{\gamma_{R2}}$ {[}panel (b), unnormalized{]}.
These zero-modes correspond to the spatially isolated PMM states $\gamma_{L1}$
and $\gamma_{R2}$, localized at the left and right QDs, respectively.{{}
}The satellite peaks originate from the \textit{bonding} ($-2\Delta\approx-24\text{{\ensuremath{\mu}eV}}$)
and \textit{antibonding} ($+2\Delta\approx+24\text{{\ensuremath{\mu}eV}}$)
hybridized states of the $\gamma_{L2}-\gamma_{R1}$ dimer formed.
Consequently, $\pi\Gamma{\cal {A}}_{\gamma_{L2}}(0)=\pi\Gamma{\cal {A}}_{\gamma_{R1}}(0)=0$.
Panel (c) exhibits the density of states (DOS) $\pi\Gamma{\cal {A}}_{d_{L\uparrow}d_{L\uparrow}^{\dagger}}(0)=\frac{1}{2}$,
characteristic of the half-electron state associated with $\gamma_{L1}$.
(d)-(f) Introducing a finite potential asymmetry ($\mu_{L}\protect\neq0$)
induces PMM spillover, reducing $\pi\Gamma{\cal {A}}_{\gamma_{L1}}(0)<1$
while activating $\pi\Gamma{\cal {A}}_{\gamma_{R1}}(0)\protect\neq0,$
a direct consequence of the $\gamma_{L1}-\gamma_{L2}-\gamma_{R1}$
trimer formation. The persistent zero-mode in $\pi\Gamma{\cal {A}}_{\gamma_{L1}}$
and $\pi\Gamma{\cal {A}}_{\gamma_{R1}}$ reflects a \textit{nonbonding}
state. Panel (f) shows the PMM spillover via $\pi\Gamma{\cal {A}}_{d_{L\uparrow}d_{L\uparrow}^{\dagger}}(0)<\frac{1}{2}$,
accompanied by a peak at $\mu_{L}=10$.}}
\end{figure}

\begin{figure}[!]
\centering\includegraphics[width=1\columnwidth]{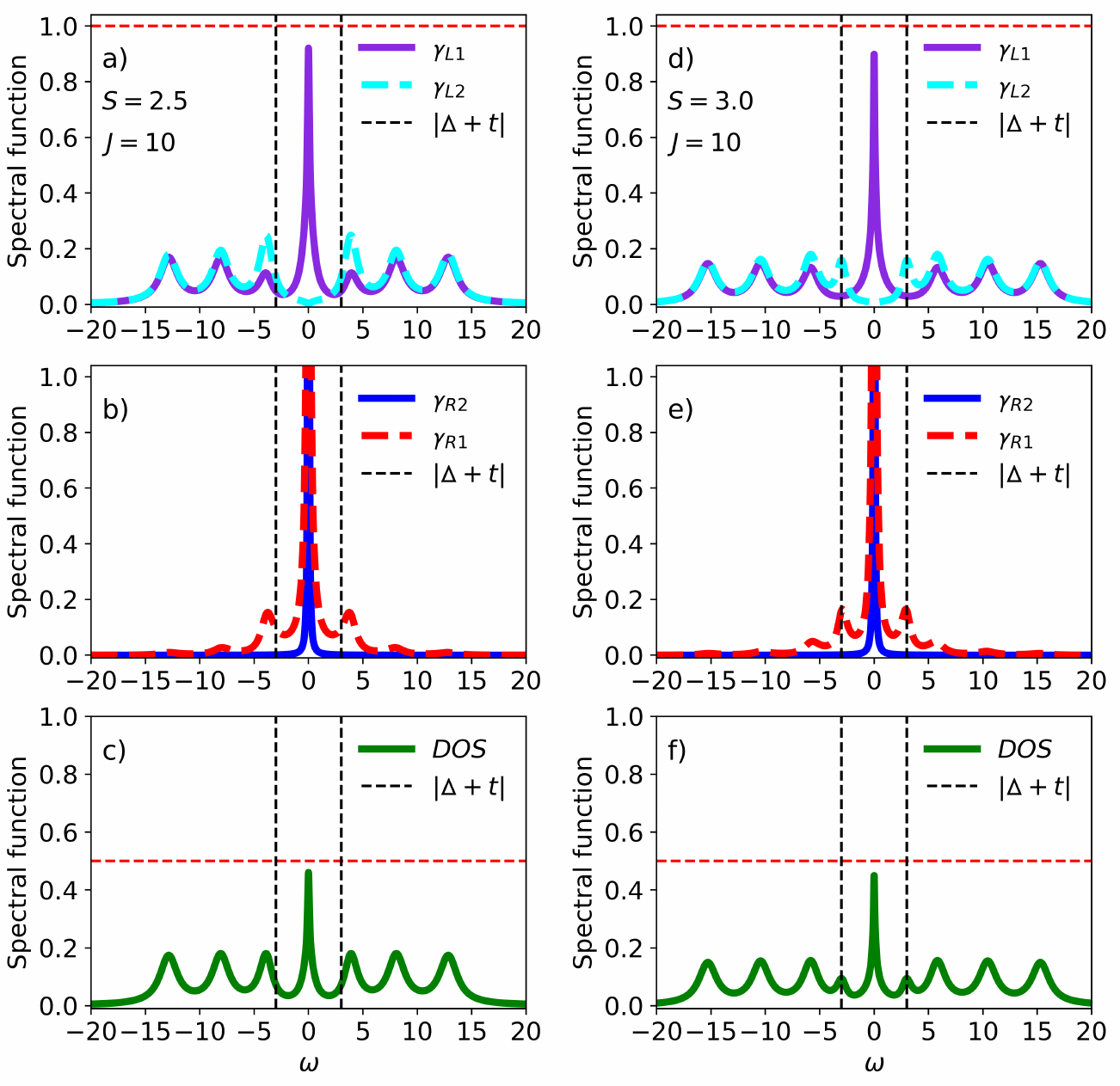} \caption{\label{Fig:Fig.3} Spin-Exchange Induced Spillover as a Spin Statistics
Probe ($t=\Delta$, $\mu_{L}=\mu_{R}=0$, and $J\protect\neq0$).
The system is governed by the trimer configuration $\gamma_{L1}-\gamma_{L2}-\gamma_{R1}$,
mediated by the exchange coupling $J$. (a)-(c) The PMM spillover
persists, evidenced by $\pi\Gamma{\cal {A}}_{\gamma_{L1}}(0)<1$ {[}panel
(a){]}, $\pi\Gamma{\cal {A}}_{\gamma_{R1}}(0)\protect\neq0$ {[}panel
(b){]}, and $\pi\Gamma{\cal {A}}_{d_{L\uparrow}d_{L\uparrow}^{\dagger}}(0)<\frac{1}{2}$
{[}panel (c){]}. For half-integer (fermionic) spin $S$, a distinct
multi-level structure with multiplicity $2S+1$ emerges around the
zero-mode, prominently visible in $\pi\Gamma{\cal {A}}_{\gamma_{L1}}$,
$\pi\Gamma{\cal {A}}_{\gamma_{L2}}$, and $\pi\Gamma{\cal {A}}_{d_{L\uparrow}d_{L\uparrow}^{\dagger}}$
due to strong spin-proximity effects. In contrast, the right QD exhibits
weaker signatures, with $\pi\Gamma{\cal {A}}_{\gamma_{R1}}$ showing
only marginal spectral weight. (d)-(f) For integer (bosonic) $S$,
the spectral structure transitions to a $2S+2$ multiplicity, particularly
pronounced in $\pi\Gamma{\cal {A}}_{\gamma_{L2}}$ and $\pi\Gamma{\cal {A}}_{d_{L\uparrow}d_{L\uparrow}^{\dagger}}$,
while $\pi\Gamma{\cal {A}}_{\gamma_{L1}}$ displays only $2S$ side
peaks. This distinction arises because the bosonic case preserves
the original \textit{bonding} ($-2\Delta\approx-24\text{{\ensuremath{\mu}eV}}$)
and \textit{antibonding} ($+2\Delta\approx+24\text{{\ensuremath{\mu}eV}}$)
states of the $\gamma_{L2}-\gamma_{R1}$ dimer ($J=0$), superimposing
them onto the $2S$-fold multiplicity. Notably, $\pi\Gamma{\cal {A}}_{\gamma_{R2}}$
remains devoid of multi-level structure, reflecting the persistent
isolation of $\gamma_{R2}$ {[}panels (b), (e){]}.}
\end{figure}

\subsection{Chemical Potential Induced Spillover}

For pedagogical clarity, we first quantify the PMM spillover induced
by electrostatic perturbation via tuning of the QD chemical potential
\cite{Flensberg_2012(Poor)}. Fig.~\ref{Fig:Fig.2} systematically
examines the PMM spillover phenomenon induced by electrostatic modulation
of the left QD chemical potential ($\mu_{L}$) \cite{Flensberg_2012(Poor)}.
We employ the Majorana spectral function $\pi\Gamma\mathcal{A}_{\gamma_{\alpha j}}$
{[}Eq.(\ref{eq:DOSMF}){]} and the electronic spectral function $\pi\Gamma\mathcal{A}_{d_{L\uparrow}d_{L\uparrow}^{\dagger}}$
{[}Eq.(\ref{eq:DOSddx}){]}, where the prefactor $\pi\Gamma$ ensures
proper normalization on the left subsystem \cite{Baranger-2011}.
We should highlight that from an experimental point of view, the differential
conductance measured at the left QD in Ref.\cite{Kouwenhoven2023},
i.e. $G_{LL},$ can be straiforwadly linked to our spectral analysis
performed for $\mathcal{A}_{d_{L\uparrow}d_{L\uparrow}^{\dagger}}.$

The \textit{sweet spot} ($t=\Delta$, $\mu_{L}=\mu_{R}=J=0$) is characterized
in Figs.\ref{Fig:Fig.2}(a)-(c) by: (i) A unit-amplitude zero-bias
anomaly in $\pi\Gamma\mathcal{A}_{\gamma_{L1}}$ {[}panel (a), $\pi\Gamma\mathcal{A}_{\gamma_{L1}}(0)=1${]},
(ii) A concomitant zero-mode in $\pi\Gamma\mathcal{A}_{\gamma_{R2}}$
{[}panel (b), $\pi\Gamma\mathcal{A}_{\gamma_{R2}}(0)\neq0${]}, manifesting
as spatially localized PMMs ($\gamma_{L1}$, $\gamma_{R2}$) at opposite
QDs. The subsidiary resonances in $\pi\Gamma\mathcal{A}_{\gamma_{L2}}$
and $\pi\Gamma\mathcal{A}_{\gamma_{R1}}$ originate from the hybridized
\textit{bonding} ($-2\Delta\approx-24\text{{\ensuremath{\mu}eV}}$)
and \textit{antibonding} ($+2\Delta\approx+24\text{{\ensuremath{\mu}eV}}$)
states of the $\gamma_{L2}-\gamma_{R1}$ dimer, which is characteristic
of the \textit{``Majorana chain regime}''. Crucially, these exhibit
complete zero-mode suppression {[}$\pi\Gamma\mathcal{A}_{\gamma_{L2}}(0)=\pi\Gamma\mathcal{A}_{\gamma_{R1}}(0)=0${]}
due to strong inter-Majorana coupling. Panel (c) reveals the hallmark
half-fermionic state through $\pi\Gamma\mathcal{A}_{d_{L\uparrow}d_{L\uparrow}^{\dagger}}(0)=\frac{1}{2}$
\cite{Baranger-2011}, reflecting the decoupled nature of $\gamma_{L1}$.

Figs.~\ref{Fig:Fig.2}(d)-(f) demonstrate the standard PMM spillover
under local gating ($\mu_{L}=10$ or $\mu_{L}\approx80\text{{\ensuremath{\mu}eV}}$
in realistic units), inducing PMM spatial redistribution. Key signatures
include: (i) Amplitude reduction in $\pi\Gamma\mathcal{A}_{\gamma_{L1}}(0)<1$,
(ii) Emergence of a zero-bias peak in $\pi\Gamma\mathcal{A}_{\gamma_{R1}}(0)\neq0$,
signifying the formation of a hybridized trimer complex ($\gamma_{L1}-\gamma_{L2}-\gamma_{R1}$).
The spectral features correspond to a \textit{nonbonding} state (zero-modes
in $\pi\Gamma\mathcal{A}_{\gamma_{L1}}$ and $\pi\Gamma\mathcal{A}_{\gamma_{R1}}$)
accompanied by \textit{bonding}/\textit{antibonding} states in $\pi\Gamma\mathcal{A}_{\gamma_{L2}}$.
The absence of a zero-mode in $\pi\Gamma\mathcal{A}_{\gamma_{L2}}$
reflects its nodal position within the trimer configuration. Fig.~\ref{Fig:Fig.2}(f)
quantitatively confirms the spillover through both the suppression
$\pi\Gamma\mathcal{A}_{d_{L\uparrow}d_{L\uparrow}^{\dagger}}(0)<\frac{1}{2}$
and the emergence of a $\mu_{L}$-resonant feature. Notably, $\pi\Gamma\mathcal{A}_{\gamma_{R2}}$
remains invariant, as $\gamma_{R2}$ maintains its isolation ($\mu_{R}=0$).

\begin{figure}[!]
\centering \includegraphics[width=1\columnwidth]{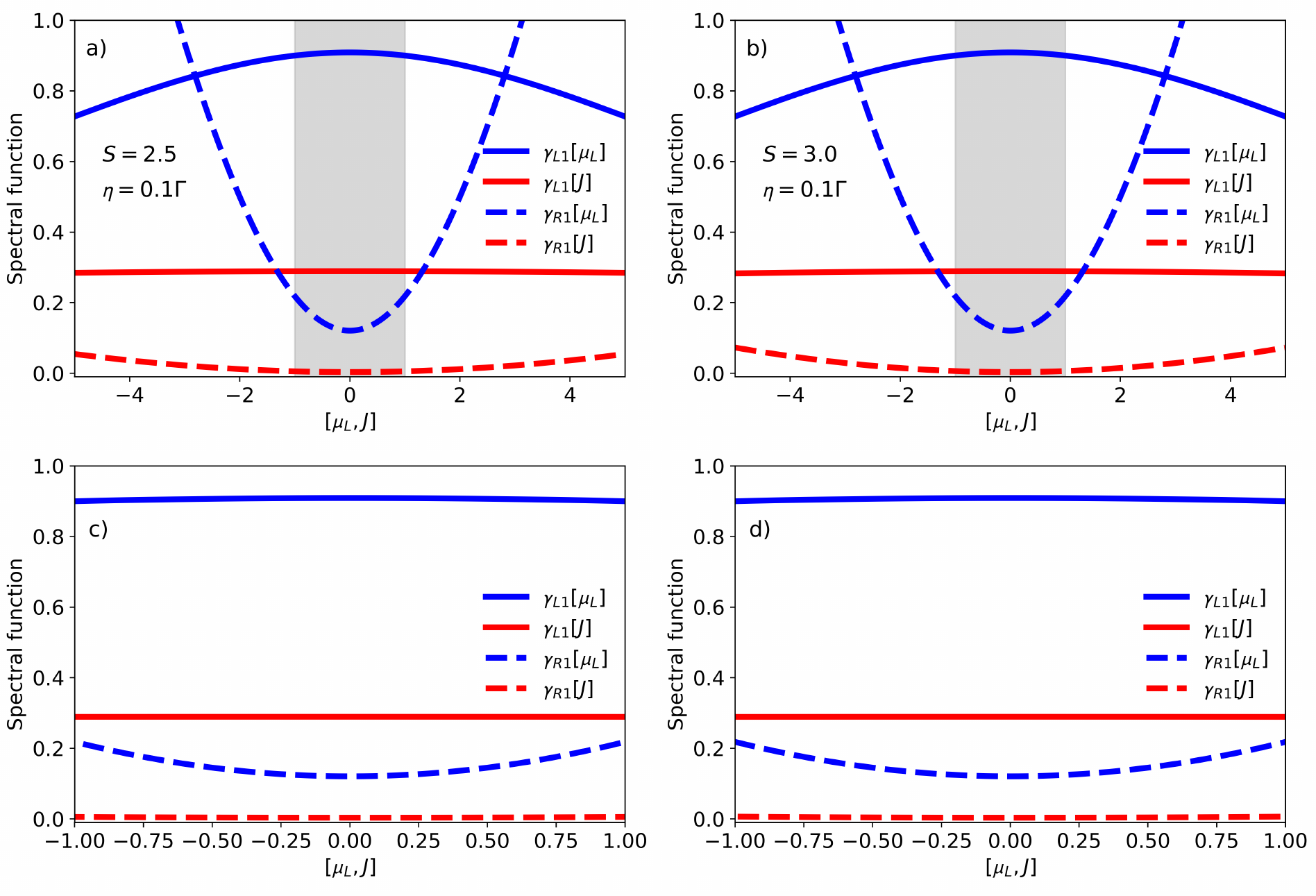} \caption{\label{Fig:Fig.4}{PMM dissipative spectral stabilization (\textquotedblleft environmentally
induced protection\textquotedblright ): (a)-(b) Spectral functions
${\cal {A}}_{\gamma_{L1}}(0)$ and ${\cal {A}}_{\gamma_{R1}}(0)$
as functions of $\mu_{L}$ and $J$. For $J$-dependent coupling at
$\mu_{L}=0$ (red curves), the presence of hybridized electronic environments
(characterized by couplings $\Gamma$ and $\eta$) enables $\gamma_{L1}$
to maintain partial Majorana dissipative spectral stabilization. This
manifests as plateaus in ${\cal {A}}_{\gamma_{L1}}(0)$ (finite) and
${\cal {A}}_{\gamma_{R1}}(0)$ (null) within the perturbation regime
$-1\lesssim J\lesssim+1$ (gray shaded region) (or $-8{\mu eV}\lesssim J\lesssim+8{\mu eV}$
in realistic units). The observed spectral stability indicates complete
suppression of PMM spillover from the left to right QD throughout
this parameter window. Remarkably, both fermionic and bosonic spin
statistics yield qualitatively similar spectral responses {[}panels
(c)-(d){]}. In contrast, for $\mu_{L}$-dependent tuning at $J=0$
(blue curves), environmental coupling fails to prevent PMM delocalization:
finite spectral weight ${\cal {A}}_{\gamma_{R1}}(0)$ emerges upon
deviation from $\mu_{L}=0$, as evidenced {[}panels (a)-(d){]}.}}
\end{figure}

\subsection{Spin-Exchange Induced Spillover}

In this subsection, we revisit the Spin-Exchange Induced Spillover
mechanism as a probe for quantum spin statistics under the conditions
$t=\Delta\approx12\text{{\ensuremath{\mu}eV}}$\cite{Kouwenhoven2023}
$\mu_{L}=\mu_{R}=0$, and $J\neq0$ \cite{Sanches_2025}. The formation
of a trimeric complex $\gamma_{L1}-\gamma_{L2}-\gamma_{R1}$ is mediated
by the exchange coupling $J$, while the characteristic signatures
of the PMM spillover persist: $\pi\Gamma\mathcal{A}_{\gamma_{L1}}(0)<1$
{[}Fig.~\ref{Fig:Fig.3}(a){]}, $\pi\Gamma\mathcal{A}_{\gamma_{R1}}(0)\neq0$
{[}Fig.~\ref{Fig:Fig.3}(b){]}, and $\pi\Gamma\mathcal{A}_{d_{L\uparrow}d_{L\uparrow}^{\dagger}}(0)<\frac{1}{2}$
{[}Fig.~\ref{Fig:Fig.3}(c){]}.

However, in contrast to chemical-potential-driven spillover, the spectral
functions exhibit a complex multi-level structure dictated by the
quantum statistics of the spin $S,$ due to $J\neq0$ (particularly,
$J=10$ or $J\approx80{\mu eV}$ in realistic units). For fermionic
spins (half-integer $S,$ here we consider $S=2.5$), the spectral
functions $\pi\Gamma\mathcal{A}_{\gamma_{L1}}$, $\pi\Gamma\mathcal{A}_{\gamma_{L2}}$,
and $\pi\Gamma\mathcal{A}_{d_{L\uparrow}d_{L\uparrow}^{\dagger}}$
display $2S+1$ distinct resonances around the zero-energy mode, reflecting
the direct exchange coupling of $S$ to the left QD. In contrast,
$\pi\Gamma\mathcal{A}_{\gamma_{R1}}$ exhibits a markedly suppressed
multi-peak structure due to weaker hybridization.

For bosonic spins (integer $S,$ here we adopt $S=3$), Figs.~\ref{Fig:Fig.3}(d)-(f)
reveal $2S+2$ satellite peaks in $\pi\Gamma\mathcal{A}_{\gamma_{L2}}$
and $\pi\Gamma\mathcal{A}_{d_{L\uparrow}d_{L\uparrow}^{\dagger}},$
whereas $\pi\Gamma\mathcal{A}_{\gamma_{L1}}$ contains only $2S$.
This disparity arises from the bosonic nature of $S$, which preserves
the \textit{bonding} ($-2\Delta$) and \textit{antibonding} ($+2\Delta$)
states of the initial dimer $\gamma_{L2}-\gamma_{R1}$ ($J=0$) even
after trimer formation ($J\neq0$), leading to $2S+2$ side resonances.
Consequently, the observation of two pinned resonances at $\omega=\pm(\Delta+t)=\pm2\Delta=\pm2t\approx\pm24\text{{\ensuremath{\mu}eV}}$
under varying $J$ permits discrimination between fermionic ($2S+1$)
and bosonic ($2S+2$) spin statistics. Noteworthy, both fine structures
appear resolved in the energy range $-20\lesssim\omega\lesssim+20$
(or $-160{\mu eV}\lesssim\omega\lesssim+160{\mu eV}$ in realistic
units), which is close to the experimental edge of visibility in experimental
setups\cite{Kouwenhoven2023}. This establishes the Spin-Exchange
Spillover as a spectroscopic tool for quantum spin characterization.
Notably, since $\gamma_{R2}$ remains uncoupled, $\pi\Gamma\mathcal{A}_{\gamma_{R2}}$
lacks any multi-level structure {[}Figs.~\ref{Fig:Fig.3}(b), (e){]}.

An alternative way of understanding that the PMM spillover mechanism
depends on the quantum spin statistics of $S$ consists in starting
the analysis of $\pi\Gamma\mathcal{A}_{d_{L\uparrow}d_{L\uparrow}^{\dagger}}$
when the CAR is absent, as pointed out by some of us in Ref.\cite{Sanches_2025}.
To perceive such, note that for $\Delta=0$ in Eq.(\ref{eq:H}) the
Kitaev dimer simplifies to an ordinary molecule with a spectrum spanning
$2\times(2S+1)$ resonant states, where the factor $2$ is reminiscent
from the \textit{bonding} and \textit{antibonding} molecular states
in the pristine condition $J=0.$ In such a scenario, the aforementioned
multiplicity is spin statistics independent. However, the distinction
between fermions and bosons for $\Delta\neq0$ arises from the mirror
symmetry of $\pi\Gamma\mathcal{A}_{d_{L\uparrow}d_{L\uparrow}^{\dagger}}$
around $\omega=|t|.$ Particularly for bosonic $S,$ we call attention
that $2S+1$ is an odd quantity, which means that for $\Delta=0$
the magnetic state $m=0$ corresponds to $\omega=|t|,$ being flanked
by $S$ states and giving rise to the $2\times(2S+1)$ fine structure,
as expected. Upon increasing $\Delta,$ the resonant state $|t|$
splits into $|\Delta-t|$ and $|\Delta+t|,$ which modifies the multiplicity
from $2\times(2S+1)$ to $2\times(2S+2).$ In case of a fermionic
$S,$ $2S+1$ is even and the magnetic state $m=0$ does not exist,
as well as the splitting due to $\Delta$ at $\omega=|t|,$ once such
a state is absent there, thus preserving the multiplicity $2\times(2S+1).$

Therefore, in the \textit{``Majorana chain regime''} ($t=\Delta$),
half of the fine structure $2\times(2S+1)$ ($2\times(2S+2)$) for
fermions (bosons) merges as a zero-mode state, while the remaining
part $2S+1$ ($2S+2$) is still resolved in $\pi\Gamma\mathcal{A}_{d_{L\uparrow}d_{L\uparrow}^{\dagger}}$
\cite{Sanches_2025}. Additionally, it is worth mentioning that although
the magnetic state $m=0$ is linked to the absence of $J,$ we count
different spectral lines in $\pi\Gamma\mathcal{A}_{d_{L\uparrow}d_{L\uparrow}^{\dagger}}$
dependent upon the $S$ magnitude and, consequently, the here reported
PMM spillover can be still considered a spin statistics effect.

In order to explore the PMM dissipative spectral stabilization due
to the couplings $\Gamma_{\alpha}=\delta_{\alpha L}\Gamma+\eta$ ($\alpha=L(R)$
for left (right) QD), a detailed analysis of the spectral profiles
${\cal {A}}_{\gamma_{L1}}(0)$ and ${\cal {A}}_{\gamma_{R1}}(0)$
as functions of $\mu_{L}$ (with $J=0$) or $J$ (with $\mu_{L}=0$)
is presented in Figs. \ref{Fig:Fig.4}(a)-(b).{{}
}At the designated \textit{sweet spot}, the $\gamma_{L2}-\gamma_{R1}$
dimer hybridizes into \textit{bonding} and \textit{antibonding} states,
precluding a zero-mode. Deviation from this point, induced by tuning
$\mu_{L}$ {[}Fig.\ref{Fig:Fig.2}{]} or $J$ {[}Fig.\ref{Fig:Fig.3}{]},
reorganizes the system into a $\gamma_{L1}-\gamma_{L2}-\gamma_{R1}$
trimer configuration. This reorganization results in the emergence
of a \textit{nonbonding} zero-mode localized on $\gamma_{R1}$, a
consequence of the spillover from $\gamma_{L1}$, thereby completing
a trivial pair of MBSs on the right QD ($\gamma_{R1}$ and $\gamma_{R2}$).

\begin{figure}[!]
\centering \includegraphics[width=1\columnwidth]{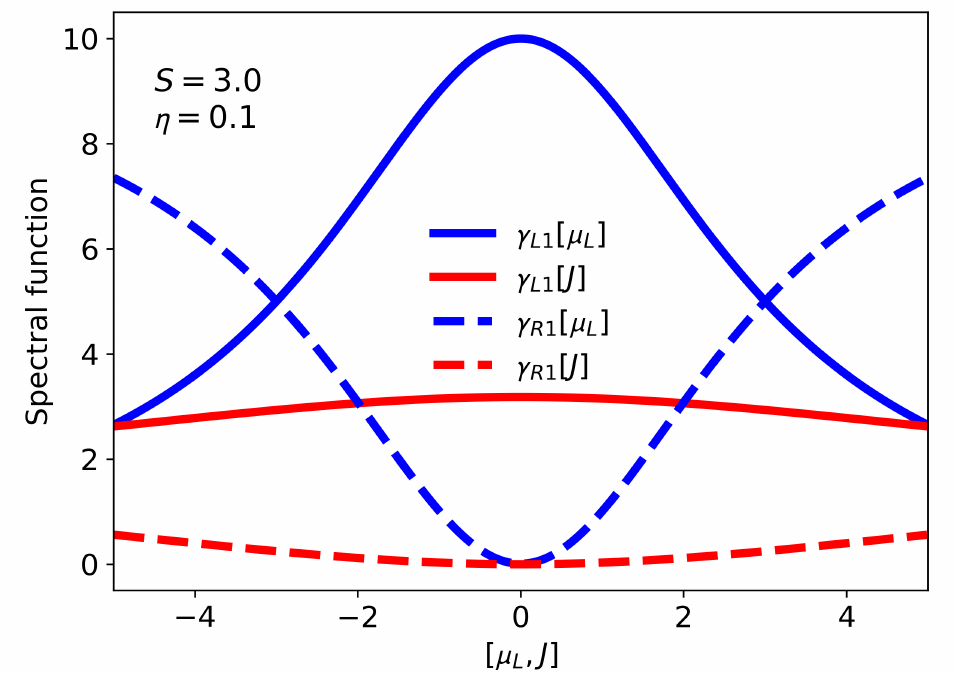} \caption{\label{Fig:Fig.5}{Spectral functions ${\cal {A}}_{\gamma_{L1}}(0)$
and ${\cal {A}}_{\gamma_{R1}}(0)$ in terms of of $\mu_{L}$ and $J:$
regardless the spin $S$ and perturbation, if the coupling $\Gamma$
is turned-off, the dissipative spectral stabilization PMM picture
breaks down. Solely at $\mu_{L}=J=0,$ the spectral weight ${\cal {A}}_{\gamma_{R1}}(0)=0.$}}
\end{figure}

\begin{figure*}[ht]
\centering\includegraphics[width=0.99\textwidth,height=0.35\textheight]{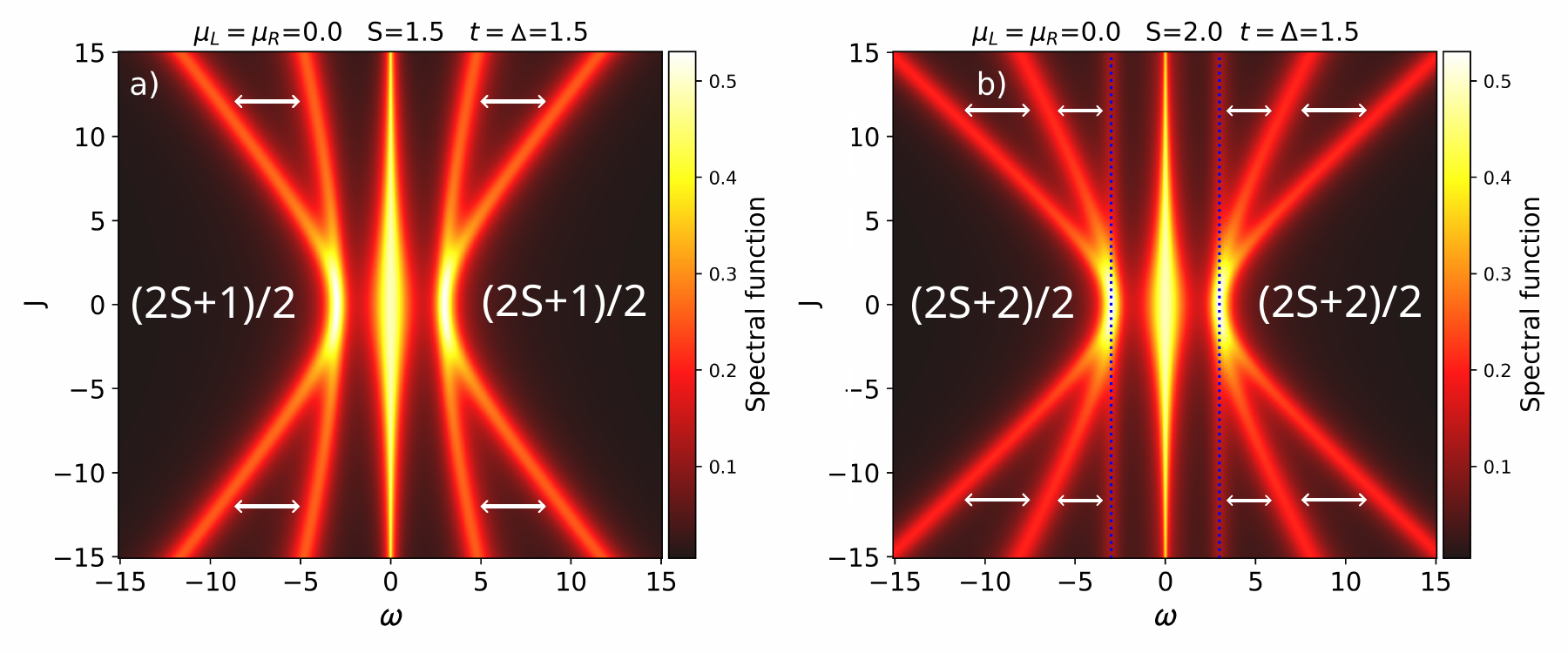}\caption{\label{Fig:Fig.6} {Spin-Exchange Induced Spillover as a Spin Statistics
Probe revealed by $\pi\Gamma{\cal {A}}_{d_{L\uparrow}d_{L\uparrow}^{\dagger}}$
spanned by $J$ and $\omega.$ These pattern spectral profiles are
the counterparts of those obtained experimentally in Refs.\cite{Kouwenhoven2023,MinimalModel2},
except that the perturbation on the left QD is ascribed to the exchange
coupling instead of the chemical potential. (a) The fermionic spin
$S$ introduces a multi-level energy structure given by the $2S+1$
multiplicity. (b) The bosonic $S$ alters such a splitting to $2S+2,$
being the inner levels (those marked by the dashed blue lines) constantly
resilient to $J$ variations characterized by peaks pinned at $\omega=\pm(\Delta+t)=\pm2\Delta=\pm2t\approx\pm24\text{{\ensuremath{\mu}eV}}.$
For fermions, such a pinning is entirely absent and the inner levels
bend upon varying the exchange coupling $J.$}}
\end{figure*}

To mitigate this spillover effect, we herein demonstrate a mechanism
for partial dissipative spectral stabilization of the proximal PMM,
which we call by \textquotedblleft environmentally induced protection\textquotedblright{}
with its genesis in the competition between the direct exchange coupling
$J$ and a multi-terminal environment defined by $\Gamma_{\alpha}=\delta_{\alpha L}\Gamma+\eta.$
 This result finds a parallel in analyses of non-Hermitian two-site
Kitaev chains \cite{NonHermitian1}. We first consider the case of
a $J$-dependent coupling at $\mu_{L}=0$ (red curves). As evidenced
in Fig. \ref{Fig:Fig.4}, partial dissipative spectral stabilization
 (PMM stability) is manifested as plateaus in the spectral functions,
where ${\cal {A}}_{\gamma_{L1}}(0)$ remains finite while ${\cal {A}}_{\gamma_{R1}}(0)$
is suppressed to zero, particularly within the interval $-1\lesssim J\lesssim+1$
(gray shaded regions) (or $-8{\mu eV}\lesssim J\lesssim+8{\mu eV}$
in realistic units). This spectral stability indicates a complete
suppression of the PMM spillover from the left to the right QD for
$|J|\lesssim1$. Notably, this behavior is largely independent of
the quantum statistics, as both fermionic and bosonic cases exhibit
close agreement {[}Figs.\ref{Fig:Fig.4}(c)-(d){]}. Conversely, for
a $\mu_{L}$-dependent tuning at $J=0$ (blue curves), the environmental
coupling is insufficient to prevent spillover. A finite spectral weight
${\cal {A}}_{\gamma_{R1}}(0)$ emerges upon even slight deviation
from $\mu_{L}=0$, as shown across all panels of Fig. \ref{Fig:Fig.4}.

Finally, upon reconfiguring the system to a symmetric coupling regime
$\Gamma_{\alpha}=\eta$ (with parameters in arbitrary units), as shown
in Fig. \ref{Fig:Fig.5}, the PMM stability is entirely eradicated,
irrespective of the statistical nature of $S$ and $\mu_{L}$ amplitude.

Interestingly enough, we should distinguish the \textquotedblleft environmentally
induced protection\textquotedblright{} here reported from the topological
protection for the Majorana based parity-qubit, which is expected
to be verified in QD arrays mimicking long Kitaev wires. To address
quantitatively the issue, we should recall the so-called qubit dephasing
time $T_{2}^{*}$ and rate $\frac{1}{T_{2}^{*}}\sim\frac{\delta\varepsilon}{\hbar},$
where the latter is the rate at which the qubit loses its quantum
coherence due to energy fluctuations $\delta\varepsilon$ arising
from the couplings to the environment (see the supplementary information
in Ref. \cite{Dephasing}). For the Kitaev model with $t=\Delta,$
the energy fluctuation arises from the energy splitting of the MBS
zero-mode when the simultaneous shift of all the on-site chemical
potentials $\mu$ is performed, thus leading to $\delta\varepsilon=2\Delta\exp[-N\ln(\frac{2\Delta}{\mu})]$
(supplementary information in Refs. \cite{MinimalModel3,Dephasing}).
This expression shows that the splitting decreases exponentially fast
with the increasing number of sites $N$ of the Kitaev wire, obeying
the rate $\ln(\frac{2\Delta}{\mu})$. As aftermath, $\frac{1}{T_{2}^{*}}$
decreases with increasing $N.$ Particularly for the Kitaev dimer,
we obtain the so-called ``quadratic protection'' $\delta\varepsilon=\frac{\mu^{2}}{2\Delta}$
\cite{Flensberg_2012(Poor)}, which should evolve to the topological
type for $N\rightarrow\infty$ \cite{Kouwenhoven2023,MinimalModel3}.
We call attention that in Ref.\cite{NonHermitian1} for the non-Hermitian
description of the Kitaev dimer, such a ``quadratic protection''
was reexamined in the presence of asymmetric metallic leads. Here,
we do not explore these features, once the right QD is not exchange
coupled to a second quantum spin. It means that local perturbations
via $J$ at both the QDs were not performed and, consequently, the
``quadratic protection'' $\delta\varepsilon\propto J^{2}$ cannot
be verified. Therefore, the current \textquotedblleft environmentally
induced protection\textquotedblright{} does not protect the Majorana
based parity-qubit.

In Fig.~\ref{Fig:Fig.6}, we conclude our analysis by presenting
the Spin-Exchange Induced Spillover Effect, characterized by $\pi\Gamma{\cal {A}}_{d_{L\uparrow}d_{L\uparrow}^{\dagger}}$
as a function of the exchange coupling $J$ and frequency $\omega$.
Without loss of generality, we change the spin to $S=1.5$ and $S=2$
for fermionic and bosonic cases, respectively. We emphasize that the
observed spectral profiles are the direct theoretical counterparts
to experimental findings reported in Refs.~\cite{Kouwenhoven2023,MinimalModel2},
with the distinction that the perturbation applied to the left QD
is the exchange coupling $J$ rather than the chemical potential.
For a fermionic spin $S$, the QD energy level structure exhibits
a multiplicity of $2S+1$ {[}Fig.~\ref{Fig:Fig.6}(a){]}, whereas
a bosonic spin $S$ results in a spectral pattern with $2S+2$ distinct
features {[}Fig.~\ref{Fig:Fig.6}(b){]}. Notably, in the bosonic
case, the inner resonances (indicated by blue dashed lines) are pinned
at frequencies $\omega=\pm(\Delta+t)=\pm2\Delta=\pm2t\approx24{\mu eV}.$
This pinning is entirely absent in the fermionic scenario, where the
inner levels exhibit dispersion as a function of $J$. Consequently,
the observation of pinned inner resonances provides a spectroscopic
signature for identifying bosonic spin character via the $2S+2$ rule,
while its absence indicates fermionic behavior, requiring application
of the $2S+1$ formula for spin determination.

\section{Conclusions}

We present a theoretical reexamination of the two-site Kitaev chain,
focusing on the hybridization dynamics of PMM modes under spin-exchange
interactions. The system exhibits a characteristic spillover effect
when coupled to a quantum spin $S$ through an exchange interaction
$J$, as recently characterized in \cite{Sanches_2025}. This spin-exchange-mediated
hybridization manifests spectroscopically through the emergence of
$2S+2$ ($2S+1$) discrete subgap states - excluding the zero-bias
anomaly - in the adjacent QD density of states, providing a direct
signature of the spin's bosonic (fermionic) quantum statistics. Remarkably,
our analysis reveals an environmental dissipative spectral stabilization:
when the minimal Kitaev chain is coupled to multiple reservoirs, the
PMM states demonstrate enhanced stability against moderate fluctuations
in the exchange coupling strength. This finding suggests that engineered
environmental coupling can provide partial localization of Majorana
modes. The combined effects of spin-exchange interactions and multi-terminal
coupling in this minimal platform yield two significant applications:
(i) A sensitive probe for quantum spin statistics through spectroscopic
measurement of the characteristic level splitting; (ii) A potentially
dissipative spectral stabilization of PMMs via the \textquotedblleft environmentally
induced protection\textquotedblright .

\section{Acknowledgments}

We thank the Brazilian funding agencies CNPq (311980/2021-0, 303896/2025-6,
308695/2021-6 and 300326/2025-4), the São Paulo Research Foundation
(FAPESP; Grant No. 2023/13467-6), Coordenação de Aperfeiçoamento de
Pessoal de Nível Superior - Brasil (CAPES) -- Finance Code 001 and
Foundation for Support of Research in the State of Rio de Janeiro
(FAPERJ Process Numbers E-26/211.605/2021 and E-26/202.421/2024). LSR acknowledges
the Polish National Science Centre, project No 2023/51/B/ST3/03025, and the Icelandic
Research Fund (Rann\'is), grant No. 239552-051.

\appendix

\section{Evaluation of the left QD normal GF}

In this appendix, we particular provide details for the evaluation
of the GF $\langle\langle d_{L\uparrow}Y^{mm};d_{L\uparrow}^{\dagger}\rangle\rangle,$
where $Y^{mm}=\left|m\right\rangle \left\langle m\right|$ and $m\in\{-S,-S+1,\dots,S-1,S\}.$
Our intend consists in offering a GF tutorial, which can be easily
extended to all the system GFs comprised by Eqs.(\ref{eq:GF1})-(\ref{eq:GF4}).
Thereby, we begin by applying the Eq.(\ref{eq:EOM}), i.e., the EOM
approach to $\langle\langle d_{L\uparrow}Y^{mm};d_{L\uparrow}^{\dagger}\rangle\rangle$
and find
\begin{eqnarray}
(\omega^{+}-\mu_{L}-\frac{Jm}{2}+i\Gamma)\langle\langle d_{L\uparrow}Y^{mm};d_{L\uparrow}^{\dagger}\rangle\rangle & = & \frac{1}{2S+1}\nonumber \\
-\Delta\langle\langle d_{R\uparrow}^{\dagger}Y^{mm};d_{L\uparrow}^{\dagger}\rangle\rangle-t\langle\langle d_{R\uparrow}Y^{mm};d_{L\uparrow}^{\dagger}\rangle\rangle & ,\label{eq:FG1}
\end{eqnarray}
where $\left\langle \left|m\right\rangle \left\langle m\right|\right\rangle =\frac{1}{2S+1}$
gives the thermal average over the spin states, $\omega^{+}=\omega+i\eta,$
$\langle\langle d_{R\uparrow}^{\dagger}Y^{mm};d_{L\uparrow}^{\dagger}\rangle\rangle$
and $\langle\langle d_{R\uparrow}Y^{mm};d_{L\uparrow}^{\dagger}\rangle\rangle$
emerge as normal and anomalous crossed GFs, respectively.

In order to continue such an evaluation, we repeatedly use the EOM
over the unknown GFs, thus giving rise to
\begin{eqnarray}
\langle\langle d_{R\uparrow}^{\dagger}Y^{mm};d_{L\uparrow}^{\dagger}\rangle\rangle & = & \frac{t}{(\omega^{+}+\mu_{R})}\langle\langle d_{L\uparrow}^{\dagger}Y^{mm};d_{L\uparrow}^{\dagger}\rangle\rangle\nonumber \\
 & - & \frac{\Delta}{(\omega^{+}+\mu_{R})}\langle\langle d_{L\uparrow}Y^{mm};d_{L\uparrow}^{\dagger}\rangle\rangle\nonumber \\
\label{eq:FG2}
\end{eqnarray}
and
\begin{eqnarray}
\langle\langle d_{R\uparrow}Y^{mm};d_{L\uparrow}^{\dagger}\rangle\rangle & = & \frac{\Delta}{(\omega^{+}-\mu_{R})}\langle\langle d_{L\uparrow}^{\dagger}Y^{mm};d_{L\uparrow}^{\dagger}\rangle\rangle\nonumber \\
 & - & \frac{t}{(\omega^{+}-\mu_{R})}\langle\langle d_{L\uparrow}Y^{mm};d_{L\uparrow}^{\dagger}\rangle\rangle.\nonumber \\
\label{eq:FG3}
\end{eqnarray}
Based on the Eqs.(\ref{eq:FG2}) and (\ref{eq:FG3}), the GF $\langle\langle d_{L\uparrow}Y^{mm};d_{L\uparrow}^{\dagger}\rangle\rangle$
of Eq.(\ref{eq:FG1}) becomes
\begin{eqnarray}
(\omega^{+}-\mu_{L}-\frac{Jm}{2}+i\Gamma)\langle\langle d_{L\uparrow}Y^{mm};d_{L\uparrow}^{\dagger}\rangle\rangle & =\nonumber \\
-2t\Delta\langle\langle d_{L\uparrow}^{\dagger}Y^{mm};d_{L\uparrow}^{\dagger}\rangle\rangle\left[\frac{\omega^{+}}{\omega^{2}-\mu_{R}^{2}+2i\omega\eta-\eta^{2}}\right]\nonumber \\
+\langle\langle d_{L\uparrow}Y^{mm};d_{L\uparrow}^{\dagger}\rangle\rangle\left[\frac{\omega^{+}(\Delta^{2}+t^{2})+\mu_{R}(t^{2}-\Delta^{2})}{\omega^{2}-\mu_{R}^{2}+2i\omega\eta-\eta^{2}}\right]\nonumber \\
+\frac{1}{2S+1} & ,\nonumber \\
\label{eq:FG4}
\end{eqnarray}
where we define the terms
\begin{eqnarray}
\tilde{K}_{R}^{+} & = & \frac{\omega^{+}(\Delta^{2}+t^{2})+\mu_{R}(t^{2}-\Delta^{2})}{\omega^{2}-\mu_{R}^{2}+2i\omega\eta-\eta^{2}}\label{eq:F1}
\end{eqnarray}
and
\begin{eqnarray}
K_{R} & = & \frac{\omega^{+}}{\omega^{2}-\mu_{R}^{2}+2i\omega\eta-\eta^{2}}\label{eq:F2}
\end{eqnarray}
to wrap up Eq.(\ref{eq:FG4}) as
\begin{eqnarray}
(\omega^{+}-\mu_{L}-\frac{Jm}{2}+i\Gamma-\tilde{K}_{R}^{+})\langle\langle d_{L\uparrow}Y^{mm};d_{L\uparrow}^{\dagger}\rangle\rangle & = & \frac{1}{2S+1}\nonumber \\
-2t\Delta K_{R}\langle\langle d_{L\uparrow}^{\dagger}Y^{mm};d_{L\uparrow}^{\dagger}\rangle\rangle & ,\label{eq:FG5}
\end{eqnarray}
here expressed in terms of the local anomalous GF $\langle\langle d_{L\uparrow}^{\dagger}Y^{mm};d_{L\uparrow}^{\dagger}\rangle\rangle,$
which by means of the EOM turns into
\begin{eqnarray}
(\omega^{+}+\mu_{L}+\frac{Jm}{2}+i\Gamma)\langle\langle d_{L\uparrow}^{\dagger}Y^{mm};d_{L\uparrow}^{\dagger}\rangle\rangle & =\nonumber \\
+t\langle\langle d_{R\uparrow}^{\dagger}Y^{mm};d_{L\uparrow}^{\dagger}\rangle\rangle+\Delta\langle\langle d_{R\uparrow}Y^{mm};d_{L\uparrow}^{\dagger}\rangle\rangle & .\label{eq:FG6}
\end{eqnarray}
After inserting Eqs.(\ref{eq:FG2}) and (\ref{eq:FG3}) into Eq.(\ref{eq:FG6}),
we obtain
\begin{eqnarray}
(\omega^{+}+\mu_{L}+\frac{Jm}{2}+i\Gamma)\langle\langle d_{L\uparrow}^{\dagger}Y^{mm};d_{L\uparrow}^{\dagger}\rangle\rangle & =\nonumber \\
+\langle\langle d_{L\uparrow}^{\dagger}Y^{mm};d_{L\uparrow}^{\dagger}\rangle\rangle\left[\frac{\omega^{+}(t^{2}+\Delta^{2})-\mu_{R}(t^{2}-\Delta^{2})}{\omega^{2}-\mu_{R}^{2}+2\omega i\eta-\eta^{2}}\right]\nonumber \\
-2t\Delta\langle\langle d_{L\uparrow}Y^{mm};d_{L\uparrow}^{\dagger}\rangle\rangle\left[\frac{\omega^{+}}{\omega^{2}-\mu_{R}^{2}+2\omega i\eta-\eta^{2}}\right] & ,\nonumber \\
\label{eq:FG7}
\end{eqnarray}
where we introduce the novel definition
\begin{eqnarray}
\tilde{K}_{R}^{-} & = & \frac{\omega^{+}(t^{2}+\Delta^{2})-\mu_{R}(t^{2}-\Delta^{2})}{\omega^{2}-\mu_{R}^{2}+2\omega i\eta-\eta^{2}},\label{eq:F3}
\end{eqnarray}
resulting in
\begin{eqnarray}
(\omega^{+}+\mu_{L}+\frac{Jm}{2}+i\Gamma-\tilde{K}_{R}^{-})\langle\langle d_{L\uparrow}^{\dagger}Y^{mm};d_{L\uparrow}^{\dagger}\rangle\rangle & = & -2t\Delta K_{R}\nonumber \\
\times\langle\langle d_{L\uparrow}Y^{mm};d_{L\uparrow}^{\dagger}\rangle\rangle\label{eq:FG8}
\end{eqnarray}
and
\begin{eqnarray}
\langle\langle d_{L\uparrow}^{\dagger}Y^{mm};d_{L\uparrow}^{\dagger}\rangle\rangle & = & -\frac{2t\Delta K_{R}\langle\langle d_{L\uparrow}Y^{mm};d_{L\uparrow}^{\dagger}\rangle\rangle}{(\omega^{+}+\mu_{L}+\frac{Jm}{2}+i\Gamma-\tilde{K}_{R}^{-})}.\nonumber \\
\label{eq:FG9}
\end{eqnarray}
Thus, by substituting Eq.(\ref{eq:FG9}) into Eq.(\ref{eq:FG5}),
we determine
\begin{eqnarray}
(\omega^{+}-\mu_{L}-\frac{Jm}{2}+i\Gamma-\tilde{K}_{R}^{+})\langle\langle d_{L\uparrow}Y^{mm};d_{L\uparrow}^{\dagger}\rangle\rangle & = & \frac{1}{2S+1}\nonumber \\
+\frac{(2t\Delta K_{R})^{2}\langle\langle d_{L\uparrow}Y^{mm};d_{L\uparrow}^{\dagger}\rangle\rangle}{(\omega^{+}+\mu_{L}+\frac{Jm}{2}+i\Gamma-\tilde{K}_{R}^{-})} & .\label{eq:FG10}
\end{eqnarray}
By defining
\begin{eqnarray}
K_{L}^{+} & = & \frac{K_{R}}{(\omega^{+}+\mu_{L}+\frac{Jm}{2}+i\Gamma-\tilde{K}_{R}^{-})}\label{eq:F4}
\end{eqnarray}
and after some algebra, we get
\begin{eqnarray}
(\omega^{+}-\mu_{L}-\frac{Jm}{2}+i\Gamma-\tilde{K}_{R}^{+})\langle\langle d_{L\uparrow}Y^{mm};d_{L\uparrow}^{\dagger}\rangle\rangle & =\frac{1}{2S+1}\nonumber \\
+(2t\Delta)^{2}K_{R}K_{L}^{+}\langle\langle d_{L\uparrow}Y^{mm};d_{L\uparrow}^{\dagger}\rangle\rangle\label{eq:FG11}
\end{eqnarray}
and
\begin{eqnarray}
(\omega^{+}-\mu_{L}-\frac{Jm}{2}+i\Gamma-\Sigma_{L}^{+})\langle\langle d_{L\uparrow}Y^{mm};d_{L\uparrow}^{\dagger}\rangle\rangle & = & \frac{1}{2S+1},\nonumber \\
\label{eq:FG12}
\end{eqnarray}
where we used $\Sigma_{L}^{+}=\tilde{K}_{R}^{+}+\left(2t\Delta\right)^{2}K_{R}K_{L}^{+},$
which is a particular version of Eq.(\ref{eq:SelfAlpha}). Thereby,
\begin{eqnarray}
\langle\langle d_{L\uparrow}Y^{mm};d_{L\uparrow}^{\dagger}\rangle\rangle & = & \frac{1}{2S+1}\frac{1}{(\omega^{+}-\mu_{L}-\frac{Jm}{2}+i\Gamma-\Sigma_{L}^{+})}.\nonumber \\
\label{eq:FG13}
\end{eqnarray}
To conclude, the left QD normal GF is finally determined by $\langle\langle d_{L\uparrow};d_{L\uparrow}^{\dagger}\rangle\rangle=\sum_{m}\langle\langle d_{L\uparrow}Y^{mm};d_{L\uparrow}^{\dagger}\rangle\rangle,$
which leads to
\begin{eqnarray}
\langle\langle d_{L\uparrow};d_{L\uparrow}^{\dagger}\rangle\rangle= & \frac{1}{2S+1}\sum_{m}\frac{1}{(\omega^{+}-\mu_{L}-\frac{Jm}{2}+i\Gamma-\Sigma_{L}^{+})} & ,\nonumber \\
\label{eq:FG14}
\end{eqnarray}
being corresponding to Eq.(\ref{eq:GF1}) with $\alpha=L.$


%

\end{document}